%% file: paper.tex
\documentclass[format=acmsmall, review=false, screen]{acmart}
\setcitestyle{numbers}
\usepackage{booktabs} 

\usepackage[ruled]{algorithm2e} 

\usepackage{pdfpages}

\usepackage{booktabs} 
\usepackage{makecell}

\usepackage{enumitem}
\usepackage{makecell}
\usepackage{relsize,color}
\usepackage{microtype}
\usepackage{framed}
\usepackage{multirow}
\usepackage[normalem]{ulem}
\usepackage{mdframed}
\usepackage{fancyvrb}
\usepackage{xcolor}
\usepackage{tabularx}
\usepackage{tikz}
\usepackage{diagbox}
\usepackage{mathtools}
\usepackage{mhsetup}
\usepackage{empheq}

\usepackage{amssymb}
\colorlet{shadecolor}{lightgray}
\usepackage{todonotes}

\usepackage{wrapfig}

\usepackage{tikz-cd}
\usepackage{subcaption}
\usepackage{tabu}

\newboolean{showcomments}
\setboolean{showcomments}{true} 
\ifthenelse{\boolean{showcomments}}
{\newcommand{\nb}[2]{
  \fcolorbox{black}{yellow}{\bfseries\sffamily\scriptsize#1}
  {\sf\small$\blacktriangleright$\textit{#2}$\blacktriangleleft$}
 }
 
}
{\newcommand{\nb}[2]{}
 
}

\usepackage{color}

\definecolor{dkgreen}{rgb}{0,0.6,0}
\definecolor{gray}{rgb}{0.5,0.5,0.5}
\definecolor{mauve}{rgb}{0.58,0,0.82}
\definecolor{weborange}{RGB}{255,165,0}

\acmJournal{TOIT}

\setcopyright{acmcopyright}

\acmDOI{0000001.0000001}

\renewcommand\footnotetextcopyrightpermission[1]{} 

\begin{document}


\title[Cloud Deployment Tradeoffs for the Analysis of Spatially-Distributed Systems of IoT]{Cloud Deployment Tradeoffs for the Analysis of Spatially-Distributed Systems of Internet-of-Things}


\author{Christos Tsigkanos}
\orcid{1234-5678-9012-3456}
\affiliation{%
  \institution{Technische Universit\"at Wien}
  \city{Vienna}
  \country{Austria}}
\author{Martin Garriga}
\affiliation{%
  \institution{Universidad Nacional del Comahue}
  \city{Neuquen, Patagonia}
  \country{Argentina}
}
\author{Luciano Baresi}
\affiliation{%
  \institution{Politecnico di Milano}
  \city{Milano}
  \country{Italy}
}
\author{Carlo Ghezzi}
\affiliation{%
  \institution{Politecnico di Milano}
  \city{Milano}
  \country{Italy}}

\renewcommand\shortauthors{Tsigkanos, C. et al}

\begin{abstract}

Internet-enabled things and devices operating in the physical world are increasingly integrated in modern distributed systems, supporting functionalities that require assurances that certain critical requirements are satisfied by the overall system. We focus here on spatially-distributed Internet-of-Things systems such as smart environments, where the dynamics of spatial distribution of entities in the system is crucial to requirements satisfaction. Analysis techniques need to be in place while systems operate to ensure that requirements are fulfilled. This may be achieved by keeping a model of the system at runtime, monitoring events that lead to changes in the spatial environment, and performing analysis. This computationally-intensive runtime assurance method cannot be supported by resource-constrained devices that populate the space and must be offloaded to the cloud. However, challenges arise regarding resource allocation and cost, especially when the workload is unknown at the system's design time. As such, it may be difficult or even impossible to guarantee application service level agreements, e.g., on response times. 
To this end, we instantiate spatial verification processes, integrating them to the service layer of an IoT-cloud architecture based on microservices. We propose several cloud deployments for such an architecture for assurance of spatial requirements --- based on virtual machines, containers, and the recent Functions-as-a-Service paradigm. Then, we assess deployments' tradeoffs in terms of elasticity, performance and cost by using a workload scenario from a known dataset of taxis roaming in Beijing. We argue that the approach can be replicated in the design process of similar kinds of spatially distributed Internet-of-Things systems.

\end{abstract}

\maketitle

\input{sections/intro}

\input{sections/example}


\input{sections/space}

\input{sections/spatial-cloud}
\input{sections/deployment}

\input{sections/evaluation}

\input{sections/relatedwork}

\input{sections/conclusions}

\bibliographystyle{ACM-Reference-Format}
\bibliography{christos}

\end{document}

%% file: sections/intro.tex

\section{Introduction}

The recent evolution towards an increasingly integrated world has at its basis novel types of pervasive systems achieved through new technologies and paradigms such as the Internet-of-Things (IoT). 
Such systems feature physically distributed devices and cloud computing infrastructure alike. This emergence comes along with new types of requirements and a need for increased assurances regarding the behavior of the overall physically-distributed systems~\cite{li2015internet}, as they permeate more and more important aspects of human activity. 

IoT systems which operate within a  dynamic spatial environment are becoming ubiquitous; think of a taxi fleet within a smart city optimizing spatial distribution with respect to passenger demand, or a manufacturing floor co-habited by humans and robots.
Those represent an important class of cyber-physical systems~\cite{baheti2011cyber}, which are faced with the manifold challenges that a dynamic spatial environment brings. They demand operational management to observe, evaluate and react to a constantly changing space.  
 When the system is operational, analysis techniques situated at {\em runtime}
are an essential prerequisite to ensure that possible changes occurring in the space  --for example due to actions performed by active agents, or by the environment itself -- do not lead to requirements violations. 
Typically, this can be achieved through a MAPE approach~\cite{kephart2003vision}; by (M)onitoring the spatial environment for changes, (A)nalyzing possible requirements violations, (P)lanning required countermeasures (e.g., moving a device from one point of space to another) and then (E)xecuting such actions and  updating the shared model of space. 
We are not concerned with 
 choice and execution of appropriate counteraction measures that may be triggered to satisfy requirements, but with the requirements analysis activity itself.
 The rationale behind this is
 that verification of the system's requirements --in the form of model checking-- is (perhaps) the most computationally intensive activity, and should be appropriately and adequately supported by the system's computational facilities. In contrast, counteractions can be actuated on the devices making up the IoT system directly. 



Requirements within IoT systems often depend on the global state of the system (that is, the position of all devices in the spatial plane) and may be formally expressed into properties and verified~\cite{fse17}.
However, providing assurances at runtime over the global state, e.g.,  through model checking, is computationally-intensive, making it unfeasible or impractical for the resource-constrained devices that populate the space in IoT scenarios~\cite{GarrigaMendonca2017}. 
A device may need to evaluate a property as part of the system's business logic; a response with the property's truth value may be required to, e.g., decide whether to perform some action or movement in the space. A naive approach would (i) place and (ii) populate instances of the model of space on each device and (iii) perform verification there -- however, storage of the model within each device, communication of the global state on each device as well as the computational load that verification incurs on resource-constrained devices renders it impractical. 

Since devices making up IoT systems are Internet-enabled, offloading the analysis computation to the cloud is a viable option~\cite{jia2017qos,xu2018joint}. However, despite the general assumption of ``infinite'' resources on a cloud deployment, additional challenges arise regarding cost and resource allocation, specially when the workload is unknown at the system's design time. In these cases, it may be difficult or even impossible to guarantee application's Service Level Agreements (SLAs), e.g., on property evaluation response times. Besides, the cloud-based architecture should be elastic enough to handle such fluctuating workloads, while conciliating providers goals and client applications needs with efficient and scalable management of applications' life-cycles. To this end, \textit{microservices}~\cite{garriga2017towards} arise as the architectural weapon-of-choice, since they are small, modular, and independently deployable and scalable in an automated way. Additionally, with the rise of function-level compute instances through Function-as-a-Service (FaaS) models~\cite{MateosFaaster17,Hendrickson:2016}, the fitness of generic cloud configurations needs to be re-evaluated for these applications.  



Our approach lies within engineering of dependable systems operating in a discrete space arising from topological relations in the spatial environment, where the information abstraction of physical \emph{location} or position of entities is inherently important~\cite{tsigkanos2016architecting,Lee.TechRep.2008}. Change of such spatial position (mobility) in distributed IoT systems -- e.g., for smart city, industrial or wireless sensor networks domains~\cite{li2015internet} -- require to periodically re-evaluate  the properties of the whole system. Firstly, assurance of such requirements may be critical, so techniques and methods for engineering dependable systems are applicable, such as formal verification. Secondly, spatial verification induces particular kinds of computational-intensive workloads, which should be supported by distributed software architectures.
Finally, regarding system operation, leveraging microservices in a cloud infrastructure for spatial analysis while ensuring non-violations of SLA must be investigated.

As the cornerstone of our approach, we instantiate verification processes, integrating them within a cloud architecture for formally verifying requirements.  Such requirements predicate on the topological distribution of devices comprising an IoT system at runtime, and formal verification is adopted to provide assurances of their satisfaction. Regarding system operation, we present the tradeoffs of the alternative deployments for such an architecture as microservices. The same type of tradeoff analysis and lessons learned can be applied to other systems in the class of systems discussed, and be easily extended to others.
Specifically, the contributions of this paper are threefold. 

{\renewcommand\labelitemi{-}
\begin{itemize}[leftmargin=.2in]
\item We address a challenging application scenario within formal assurance of global requirements of spatially-distributed IoT systems. We define the underlying model of space as a generic graph structure -- a closure space~\cite{galton2003generalized} -- and express global properties of the overall system in a topological spatial logic~\cite{ciancia2014specifying}.

\item We instantiate verification processes as (FaaS-ified~\cite{MateosFaaster17}) microservices encapsulating spatial model checkers and instances of space models,
 integrating them to the service layer of the IoT-cloud architecture~\cite{li2015internet}.
 Finally, we position the microservices-based architecture into the IoT system environment, involving resource-constrained devices connected to the Internet.


\item We propose alternative cloud deployments for such a model checking microservices-based distributed system, based on Virtual Machines (VMs), containers, and FaaS. We investigate their tradeoffs in terms of performance, scalability, elasticity and cost when evaluating global spatial properties at runtime. We further consider a novel \textit{hybrid} deployment, combining VMs/Containers and FaaS. We discover whether the different cloud deployments are able to cope with given strict Service Level Agreements (SLAs) using a realistic workload scenario from a known dataset of taxis roaming in Beijing, as an representative instance of the class of systems we consider. 
\end{itemize}
}

The rest of the paper is structured as follows. Section~\ref{sec:motivation-example} presents a motivating example for spatial analysis at runtime. Section~\ref{sec:space} details our proposal for reasoning on spatial properties of topological spaces. Section~\ref{sec:spatialcloud} discusses how to perform spatial verification at runtime as an IoT-cloud architecture, based on  microservices. Section~\ref{sec:deployment} details the different deployment alternatives and for such an architecture and their tradeoffs for offloading the reasoning machinery to the cloud. Section~\ref{sec:evaluation} discusses the evaluation of our approach and the alternative deployments through a well-known dataset of taxis in Beijing. 
Section~\ref{sec:relatedwork} presents related work.
Finally, Section~\ref{sec:conclusion} concludes the paper.

%% file: sections/example.tex

\section{Motivating example: Bike-Sharing in a Smart City}
\label{sec:motivation-example}

In this section, we introduce a motivating scenario as a running example of a spatially-distributed IoT system throughout the paper. We begin with a brief description of the scenario's setting and then consider SLA goals that the system should exhibit as well as the challenges that arise.

Consider a bike-sharing scenario in a smart city, where bicycles roam in the physical space. 
As is common in such cases\footnote{E.g., mobike: \url{https://mobike.com/global/}.}, bikes --or their users-- maintain a connection to the network, rendering the system an instance of the IoT.
A key issue in bike-sharing systems, is the concentration of bikes in certain parts of the city, other parts remaining empty, a phenomenon referred to as clamping. This may occur for instance around the start or end of the  business day throughout the week, or due to sporadic events such as public concerts.
The city administrator as the bike-sharing operator, has a significant interest to avoid this, as it leads to unavailability of bikes in other parts of the city when clamping occurs elsewhere.
In our scenario, to mitigate this, the city identifies certain bike usage situations that describe optimal use of the bikes; a form of quality-of-service sub-goals that aim at avoiding undesired distribution of bikes.
To make such cases attractive to bicycle commuters, the city offers rewards to bike-sharing users if a sub-goal is successfully satisfied for their bike.
The sub-goals may reflect intuitive mitigations to the clamping problem, e.g. leaving the bikes around metro stations so that they can be readily re-used by others. However, clamping mitigations might be complex, capturing cases obtained from transportation analysis of the city.
As an example, we consider the following two sub-goals --a simple and a complex one-- of the bike-sharing system, concerning every single bike:

\begin{enumerate}[label=(G\arabic*)]  
  \item Bikes should not be located close to each other at any point. Specifically, no bike should be 
  located in a place nearby where another bike is. 
  \item Bikes should be located in places in the city from which one can reach certain important landmarks --such as the main square-- through the transportation network of bus and underground, but without traversing bridges. Additionally, no bus stops should be traversed if other bikes are located there.
  \end{enumerate}

\begin{wrapfigure}{r}{0.45\textwidth}
  \centering
  \includegraphics[width=0.5\textwidth]{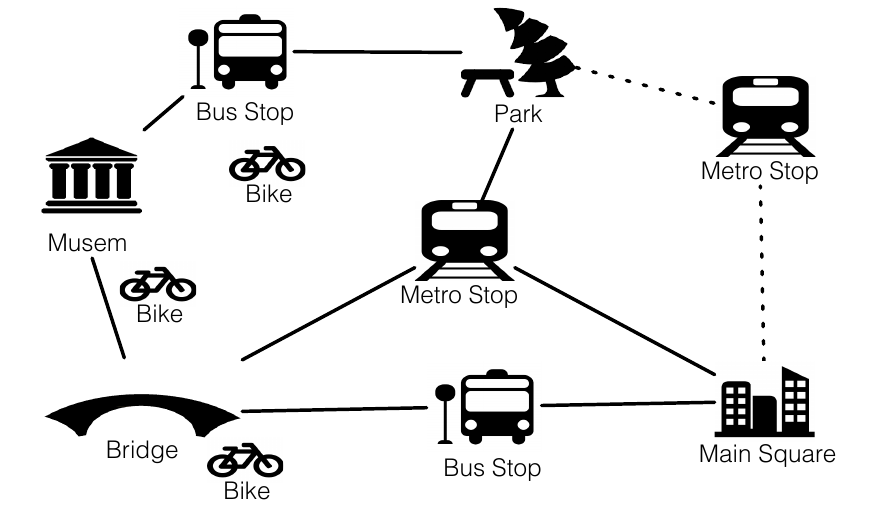}
  \caption{Topological accessibility in a city.}
  \label{fig:examplegraph}
\end{wrapfigure}

Inherent in the formulation of the scenario above is a description of how various points in the city are connected, meaning that one can use a bike to go from a point to another.
Such an \emph{accessibility} model can be leveraged from existing map data
or readily constructed by monitoring bike usage over time. The result is a graph structure, which has nodes corresponding to various points of the city and edges connecting those which are directly reachable -- a topological map of the physical reachability of the city. For our example city, such a topological graph is depicted in Figure~\ref{fig:examplegraph}, where various city landmarks such as a park, a museum and transportation stops are connected due to being directly reachable from each other by bikes.


Challenges inherent to the motivating scenario considered include both evaluating the truth values of sub-goals (G1) and (G2) for all bikes in the city at any moment, as well as architecturally supporting this analysis. 
Notice how the evaluation of sub-goals for each bike depends on the position of all other bikes in the city -- the system's global state-- rendering the sub-goals \emph{global spatial properties}~\cite{fse17}. Additionally, evaluation of such global properties is challenging because of computational and scalability concerns, requiring a dedicated architecture for the overall IoT system. 
A na{\"i}ve approach would be to perform computation upon the devices constituting the IoT system themselves (assuming that they are communicated the global system state). In this case, the advantages are reduced latency and minimal persistent connectivity needs, as long as devices support the computation workload. However, the main disadvantage of this approach is the severe limitation on the computing resources of the IoT devices~\cite{song17offloading}. Even considering a top-of-the-range smartphone or tablet, SLA constraints and prohibitive battery drain make this approach unfeasible in practice~\cite{GarrigaMendonca2017} (see also Section~\ref{sec:evaluation} for workload comparisons).

%% file: sections/space.tex

\section{Reasoning on Spatially-Distributed Internet-of-Things Systems }
\label{sec:space}

In this section, we present our approach for reasoning about global spatial properties 
of (finite) models of topological space, captured by the spatial distribution of an IoT system.
These models are closure spaces~\cite{galton2003generalized}, a generalization of a standard \emph{topological} space.
Subsequently, we utilize a Spatial Logic for Closure Spaces (SLCS)~\cite{ciancia2014specifying}, an extension of 
the topological semantics of modal logics to closure spaces, to support formal reasoning. In Section~\ref{sec:space-evalmodel}, we first provide a context for evaluating spatial predicates, by showing how a spatial evaluation model may be derived.
This model is illustrated over the accessibility in the example city previously described. Given such an evaluation model, in Section~\ref{sec:spaceverif} global properties can be verified.

\subsection{Evaluation Model of Space}
\label{sec:space-evalmodel}

Our evaluation model is based on the established notion of trajectories~\cite{zheng2011computing} in geographical space, upon which many datasets and applications build. Such trajectories are often available as a domain model; for our running scenario, trajectories of bikes can be obtained e.g. by monitoring the bike-sharing system usage for a limited amount of time~\cite{adrienko2011spatial}. In the following, we demonstrate: (i) how a topological accessibility graph may be obtained from such a trajectory dataset; and (ii) how a discrete dataset consisting of traces of presences may be obtained. We obtain a topological \emph{accessibility} graph in two steps:

\begin{enumerate}
\item Obtain the points-of-interest (POIs) in the city. These are nodes of the graph, each consists of a name, various attributes and geographical coordinates of the point in geographical space, sourced from some widely available repository (e.g. Open Street Map~\cite{openstreetmap} or SPOI~\cite{cerba2016smart}).
\item Map trajectories of active entities over the POIs. If, based on a predefined distance, a trajectory passes through two points, these two points are considered \textit{accessible} from one another, and an edge between the respective nodes is created on the graph.
\end{enumerate}

 \begin{wrapfigure}{r}{0.45\textwidth}
  \centering
  \includegraphics[width=0.5\textwidth]{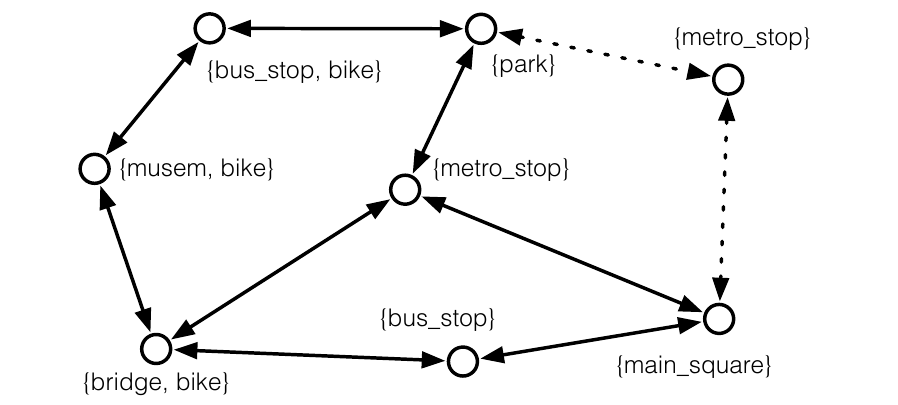}
  \caption{Propositions associated to connected points form a closure model for the example city.}
  \label{fig:closuremodel}
\end{wrapfigure}

Back to the running example, if a bike starts from the POI \emph{museum} (Figure~\ref{fig:examplegraph}), goes through the \emph{bridge} and arrives near the subsequent \emph{metro stop}, then the three are  considered (step-wise) connected, since accessibility was demonstrated by at least one trajectory. The resulting graph will have an edge linking the museum with the bridge and another linking the bridge to the metro stop.  The predefined distance aims to mitigate sensing errors in the data, as well as resolution within the scale of which POIs are defined. The result of the process is a graph capturing accessibility of POIs in geographical space. Note how time/distance may be introduced to this model by time-stamping points of a trajectory when mapping over POIs and taking into account trajectory traversal time. The additional time dimension leads to a model describing trajectories as traces of presences -- analogously, accounting distances enables quantitative reasoning.

The obtained accessibility graph bidirectionally connects POIs.  
This adjacency relation between POIs may induce a \emph{closure space}~\cite{galton2003generalized}, a mathematical model which can be used for formal reasoning with a spatial logic. A closure space is a notion originating from the field of 
mathematical topology, built upon what can be informally referred as the 
``least possible enlargement'' of a set and exhibiting certain fundamental axioms.
 Formally, a closure space is a pair $(X, \mathcal{C})$ where $X$ is a set, and the closure operator $\mathcal{C} : 2^X \rightarrow 2^X$ which assigns to each subset $A$ of $X$ its closure, such that for all $A, B \subseteq X$:
 \begin{align*}
     \mathcal{C}(\emptyset) = \emptyset;\ A \subseteq \mathcal{C}(A)\ \text{and}\ \mathcal{C}(A \cup B) = \mathcal{C}(A) \cup \mathcal{C}(B).
\end{align*}

The elements of $X$ are called the {\em points} of the closure space $(X, \mathcal{C})$.
For any subset $A \subseteq X$, we define the {\em complement} of $A$ in $X$ as $\overline{A} = X \setminus A$.
 For a closure space $(X, \mathcal{C})$, for each $A \subseteq X$, the {\em interior} $\mathcal{I}(A)$ of $A$ is the set $\overline{\mathcal{C}(\overline{A})}$.
Moreover, the {\em boundary} of a set informally refers to the set of elements which can be approached both from inside it and from outside of it. 
Formally, in a closure space $(X, \mathcal{C})$, the boundary of  $A \subseteq X$ is defined as $\mathcal{B}(A) = \mathcal{C}(A) \setminus \mathcal{I}(A)$. 
 Two more variants of boundary exist,  the {\em interior boundary} $\mathcal{B}^{-} (A) = A \setminus \mathcal{I}(A)$, and the {\em closure boundary} $\mathcal{B}^{+} (A) = \mathcal{C}(A) \setminus A$. 
Finally, as shown in~\cite{galton2003generalized,ciancia2014specifying},
every graph for which the set of edges forms a binary relation induces a
closure space, called a {\em quasi-discrete closure space}, by interpreting 
closure as the adjacency of the nodes in our accessibility graph.  
Given a set of propositions $P$, a closure model~\cite{ciancia2014specifying} is a pair $M = (X, \mathcal{C}), V$ consisting of a closure space $(X, \mathcal{C})$ and a valuation $V : P \rightarrow 2^X$ , assigning to each proposition the set of points where the proposition holds.

In Figure~\ref{fig:closuremodel}, the binary relation of \emph{accessibility} between locations (POIs or landmarks) gives rise to a closure space. Thereupon, propositions associated with each point capture the POI name as well as presence of a bike. 


  \subsection{Verification in Space}
\label{sec:spaceverif}

Building upon the previously defined fundamental operators of closure and boundary,
we proceed to briefly outline the syntax and semantics of SLCS~\cite{ciancia2014specifying}, 
a spatial logic for closure spaces. The logic will be evaluated upon models $\mathcal{M}$.
The logic features boolean operators, a ``one step'' modality turning closure into a logical operator, 
and a spatially interpreted \textit{surrounds} operator\footnote{Intuitively, closure and surrounds in a 
quasi-discrete closure space behave similarly to next and until in the temporal domain.}.
Given that $p$ ranges over a set of propositions $P$, the syntax of SLCS is defined by the following grammar:
\begin{small}
      \begin{empheq}[]{align}
    \label{syntax}
    \phi ::= p ~|~ \top ~|~ \neg \phi ~|~  \phi \wedge  \psi ~|~ \mathcal{C}\ \phi ~|~  \phi\ \mathcal{S}\  \psi.
    \end{empheq} 
\end{small}
In Formula~\ref{syntax}, $\top$ denotes true, $\neg$ is negation, $\wedge$ is conjunction, $\mathcal{C}$ is the closure operator, and $\mathcal{S}$ is the spatial surrounds operator.
Satisfaction $\mathcal{M}, x \models \phi$ of formula $\phi$ at point $x$ in model $\mathcal{M} =
((V, \mathcal{C}), v)$ is defined inductively on terms~\cite{ciancia2015experimental}.
More complex logical operators can be defined based on the fundamental 
operators of closure and spatial surrounds.
 In the following, we recall fundamental ones~\cite{ciancia2014specifying,fse17} being useful in the context of this 
paper and later used for the evaluation of our approach in 
Sec.~\ref{sec:evaluation}.
 First, we consider the intuitive notion of \emph{nearness}; points that are \emph{near} another in a model can 
be found inside the set identified by (applications of) the closure operator upon this point.
 Thus, nearness can be defined by nesting applications of the closure operator as $\mathcal{N}^n\ \phi \overset{def}= \mathcal{C}^n \phi$.
Operator $\mathcal{N}$ can be applied arbitrarily often to 
predicate about points being in a defined {\em proximity} from 
each other. For example, based on the closure model of Figure~\ref{fig:closuremodel}, $\mathcal{N} (\{ museum,bike \})$ will yield the three points on the leftmost part of the graph.

We further consider an operator $\mathcal{T}$ based on the spatial surrounds operator~\cite{ciancia2014specifying} which captures a notion of \textit{reachability}, defined as 
$\phi\ \mathcal{T}\ \psi \overset{def}= \phi \wedge \big( \neg \big((\neg\psi)\ \mathcal{S}\ (\neg (\phi \vee \psi ))\big) \big)$.
$\phi\ \mathcal{T} \psi$ is 
satisfied for a point $x$ if it satisfies $\phi$ and we can reach a point 
satisfying $\psi$ while passing only by points satisfying $\phi$. 
Then, based on $\mathcal{T}$, a more complex \textit{reach through}  operator $\mathcal{\Re}$ defined as $\phi\ \mathcal{\Re}(\psi)\  \zeta \overset{def}=\phi\ \mathcal{T}\ \big( (\psi\ \mathcal{T}\ \zeta) \wedge (\psi\ \mathcal{T}\ \phi)\big)$,
is satisfied for a point $x$ if $x$ satisfies $\phi$ and there is a sequence of points starting
from $x$, all satisfying $\psi$, reaching a target point satisfying $\zeta$.
The interested reader is referred to~\cite{ciancia2014specifying}, as we consider defining complex operators as out of scope for this paper.
\begin{small}
      \begin{empheq}[]{align}
      & \mathtt{bike}\ \mathcal{\Re}\Big(\mathtt{(!bridge \vee \big( bus\_stop \wedge bike \big) }\Big)\  \mathtt{main\_square}. \label{reach-example}
    \end{empheq} 
\end{small}
For an example illustrating the use of the derived $\mathcal{\Re}$ operator, consider the Formula~\ref{reach-example} 
which specifies the {\small $\mathsf{main\_square}$} being reachable from a point which has the proposition 
{\small $\mathsf{bike}$}. 




%% file: sections/spatial-cloud.tex

\section{Spatial Analysis at Runtime for the Internet-of-Things}
\label{sec:spatialcloud}

In this section we discuss bringing spatial verification at runtime. After outlining essential computational properties of the model checking workload, we propose an application model where the workload is deployed on the cloud. 
Our proposal architecturally consists of software services and has as its main goals the satisfaction of SLAs and the optimization of the cloud resources used.

\subsection{Spatial Verification as a Computation-Intensive Workload} 
\label{sec:spatialverifworkload}

Evaluation of spatial properties expressed in the SLCS syntax of Formula~\ref{syntax} over a closure model constitute a specific type of workload. The major computational component of this workload is the actual verification procedure -- other minor tasks such as data structure manipulation are insignificant with respect to the computational burden of explicit-state model checking. 

The model checking procedure evaluates SLCS properties over the closure model representing the current state of the system. Recall that the model consists of a graph structure (POIs and their adjacency relations, capturing accessibility in the model) as well as certain propositions that hold in points (e.g. presence of bikes). 
Following the model checking approach presented  in~\cite{ciancia2015experimental}, evaluation of a formula yields the set of points of the closure space where the formula is true. In our case, we return a truth value which indicates if the obtained set is non-empty. That is, the existence or absence of points in the closure model for which the evaluated formula is true. 
Thus, note that Formula~\ref{reach-example} is violated for the {\small $\mathsf{bike}$} in the {\small $\mathsf{museum}$} in the upper left corner of Figure~\ref{fig:closuremodel}, but satisfied for the other two bikes located in the {\small $\mathsf{bus\_stop}$} and the {\small $\mathsf{bridge}$}.

Given a spatially distributed system, representation of its discrete underlying structure as a set of points and their adjacency can be considered as virtually constant throughout its operation. With respect to the motivating example of Section~\ref{sec:motivation-example}, this corresponds to the topological structure of the city. The closure model additionally records the current positions of the bikes through propositions. Updating bike positions -- changing the propositions assignment and subsequently verifying a property such as the one of Formula~\ref{reach-example}, is performed in a near-constant time. The actual verification burden induced depends on the size of the particular closure model, as well as on the complexity of the properties checked. However, the overall operation has challenging characteristics when considered as a computational workload. Spatial model checking as presented is a very computation-intensive operation, since it requires verifying the truth value of a complex formula. However, memory requirements are near-constant, since model checking requires keeping the closure model in memory, the size of which is known beforehand (i.e., it is the model \emph{size} of the city). Such characteristics of the computational workload are exploited to design the microservices architecture presented in the following section, enabling spatial model checking to be performed at runtime while meeting strict SLAs.   

Finally, we note that this renders the verification workload fundamentally different in contrast to model checking approaches on the cloud verifying temporal logic properties~\cite{lerda1999distributed,bellettini2013distributed,brim2005assumption}. Given the above, we are not interested in partitioning or parallelizing verification computation, as our domain is spatial model checking, and the size of the space is virtually fixed per problem instantiation -- we seek to evaluate properties that otherwise components in an IoT system would compute, given the particularities of the spatial verification workloads induced.

\subsection{Service-based IoT Spatial Analysis in the Cloud}
\label{sec:iot-arch}

Given the limitations to perform computation in the devices themselves, offloading and deploying such computation to the cloud is a viable option~\cite{jia2017qos,xu2018joint}. In the cloud, computing resources are typically provided through virtualization and containerization~\cite{Quatrocchi2016discrete,leitner2016patterns}, and there is an illusion of infinite resource availability thanks to horizontal scaling. However, accessing these resources may involve multiple hops of network communication, adding prohibitive latency in the processing of client requests. In this particular case of analysis however, computation is the major cause of delays.
Thus, we propose an application model that brings together stateless components and immutable data, while stateful components are deployed separately. The proposed model prevents data consistency problems that would arise if stateful components such as databases were deployed in a distributed fashion. It also allows multiple service instances to coexist and be deployed independently without the need for state migration, which is particularly important to cope with client mobility. Indeed, as devices will freely enter and exit geographical areas, even cloud providers may see considerable variations to their aggregate demand over time. In such a scenario it may be unfeasible to predict the origin and intensity of the workload~\cite{mendonca18toit}. 


When managing IoT services in the cloud, there are two conflicting goals: i) the satisfaction of SLAs, and ii) the optimization of the resources consumed by these services in the cloud. In this paper we focus in two kinds of application requirements, namely service latency and cost; simultaneously we target efficient and scalable usage of computational resources from cloud providers -- mainly CPU, since our applications are computation intensive~\cite{Quatrocchi2016discrete}.
From the provider's perspective, an efficient and scalable allocation of its (virtualized) resources must be allocated to cope with the SLA of each provided service. To be efficient, the allocation of resources should be able to mimic the corresponding fluctuations of demand, i.e., be highly responsive. Consequently, it is important that the mechanisms governing resource allocation be aware of the actual, and potential workload~\cite{mendonca18toit}.

\begin{figure}[htpb]
	\centering
	\vspace{-.4cm}
	\includegraphics[width=.78\textwidth]{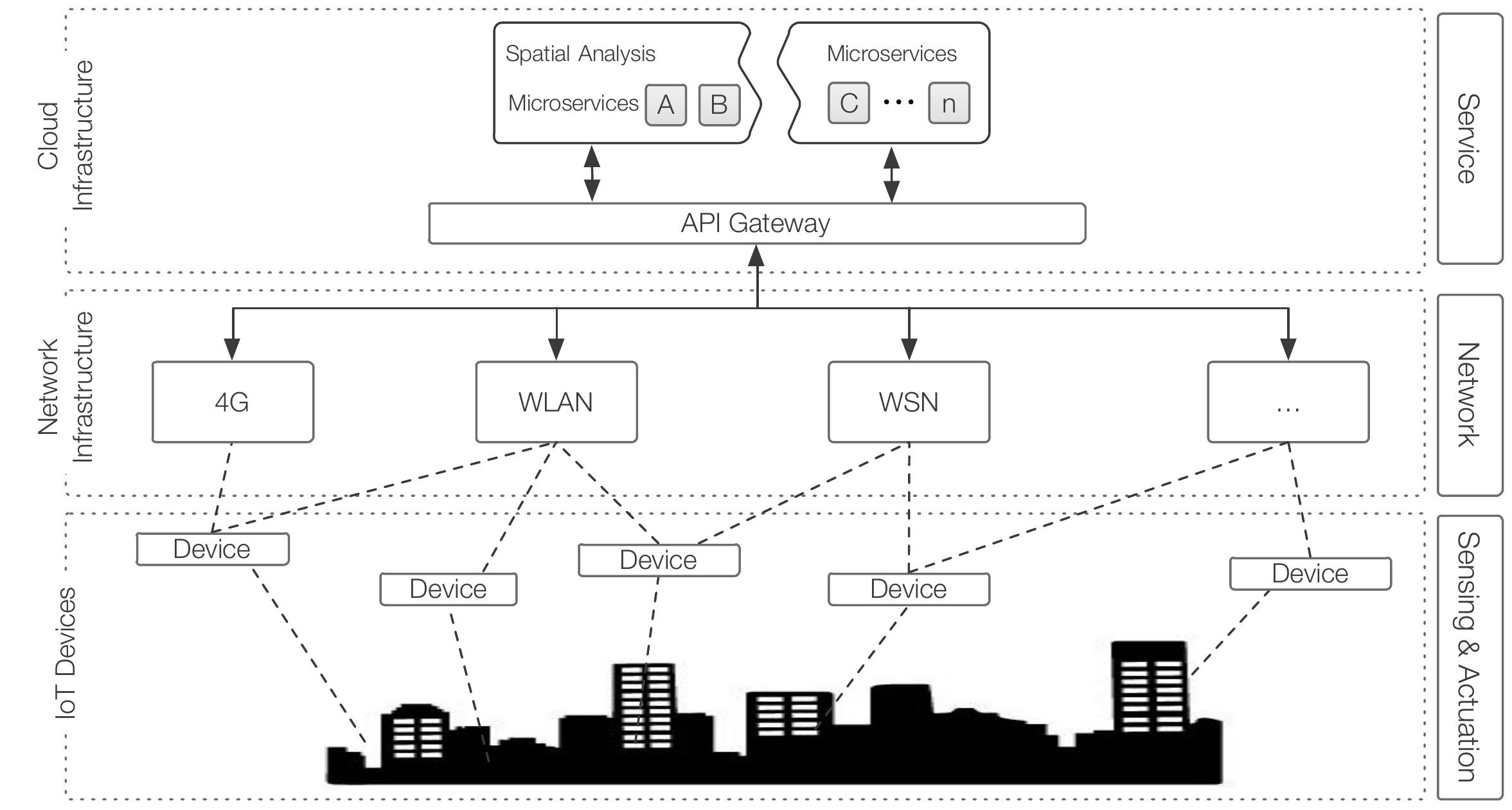}
	\vspace{-.5cm}
	\caption{IoT and microservices architecture for materializing spatial reasoning on the cloud.}
	\label{fig:arch-iot-high-level}
	\vspace{-.3cm}
\end{figure}

IoT system architectures bridge the gap between the physical and the virtual worlds, but entail multiple challenging factors. IoT architecture design involves networking, communication, and interoperability among heterogeneous devices~\cite{li2015internet}, while the overall system must exhibit scalability. These devices may be heavily resource-constrained, e.g., in terms of computational power and battery constraints, while IoT applications have to meet stringent quality of-service requirements, hindering the provision of computing and communication elements~\cite{Yu18infocom}. 
Due to the fact that things may move in the physical space and need to interact with others in real-time, an IoT architecture should be adaptive and decentralized, as well as support technology-agnostic event-based communication between heterogeneous devices serving as the fundamental architecture components. In this context, Service-Oriented Architectures (SOA) ensure interoperability among the heterogeneous devices~\cite{li2015internet} and allows to treat such a complex system as a set of well-defined simple services~\cite{baresi2017microservices}. 
Due to these advantages, SOA has been widely applied as a mainstream architecture, for example in the context of Wireless Sensors Networks (WSNs)~\cite{chi2014reconfigurable}. SOA applied to IoT provides extensibility, scalability, modularity and interoperability among heterogeneous things; in addition, the functionalities and capabilities are abstracted into a common set of services.
Figure~\ref{fig:arch-iot-high-level} provides a generic SOA view for IoT in the context of our problem, which consists of three layers~\cite{li2015internet}:
\begin{itemize}
	\item Sensing/Actuating layer, integrated with available devices, responsible for obtaining their status and acting upon decisions based on analysis outcomes.
	\item Network layer, reflecting the infrastructure that supports wireless or wired connection among things, and
	\item Service layer, which creates and manages services used by end users and applications.
\end{itemize}
	
In our approach, IoT services are indeed microservices~\cite{garriga2017towards,lewis2014microservices}, as they are small, modular, communicate with lightweight mechanisms (often through an HTTP RESTful API) and are independently deployable by fully automated machinery.
Back to our running example, a bike's location sensors (sensing layer) emit heartbeats with its coordinates, either periodically or upon certain events (e.g. locking/unlocking, or in proximity to certain landmarks or POIs). Such heartbeats should trigger the evaluation of the global model in order to determine whether the defined sub-goals hold for the current position of the bike (as well as the positions of all other bikes). 
The event travels across the network infrastructure towards the cloud (Network layer), reaching the cloud service in charge of processing it (Service Layer).
Finally, the result of the evaluation travels back to the actuator layer, to undertake the corresponding actions e.g. through the mobile device of the user (e.g., providing feedback regarding penalties or rewards). 

%% file: sections/deployment.tex
\section{Cloud Deployments for IoT Spatial Reasoning}
\label{sec:deployment}

This section details the materialization of our proposal as a 
microservices architecture for IoT (based on the SoA-IoT interplay discussed in Section~\ref{sec:iot-arch}), which allows to offload the reasoning on space presented in Section~\ref{sec:space}. 
We instantiate verification processes integrating them to the service layer of the IoT-cloud architecture~\cite{li2015internet} of a spatially-distributed IoT system involving resource-constrained devices connected to the Internet.
Applications with specific processing demands, including the computation-intensive ones discussed through this paper, have historically required special configurations such as compute- or memory-optimised virtual machine instances~\cite{MateosFaaster17}. With the recent rise of function-level compute instances through Function-as-a-Service (FaaS) models, the fitness of cloud configurations needs to be re-evaluated for these applications. Subsequently, we propose alternative cloud deployments for such a model checking  microservices-based architecture, based on Virtual Machines (VMs), containers, and FaaS, as well as a \textit{hybrid} deployment setup.


\begin{figure}[htpb]
  \vspace{-.2cm}
\centering
	\includegraphics[width=.9\textwidth]{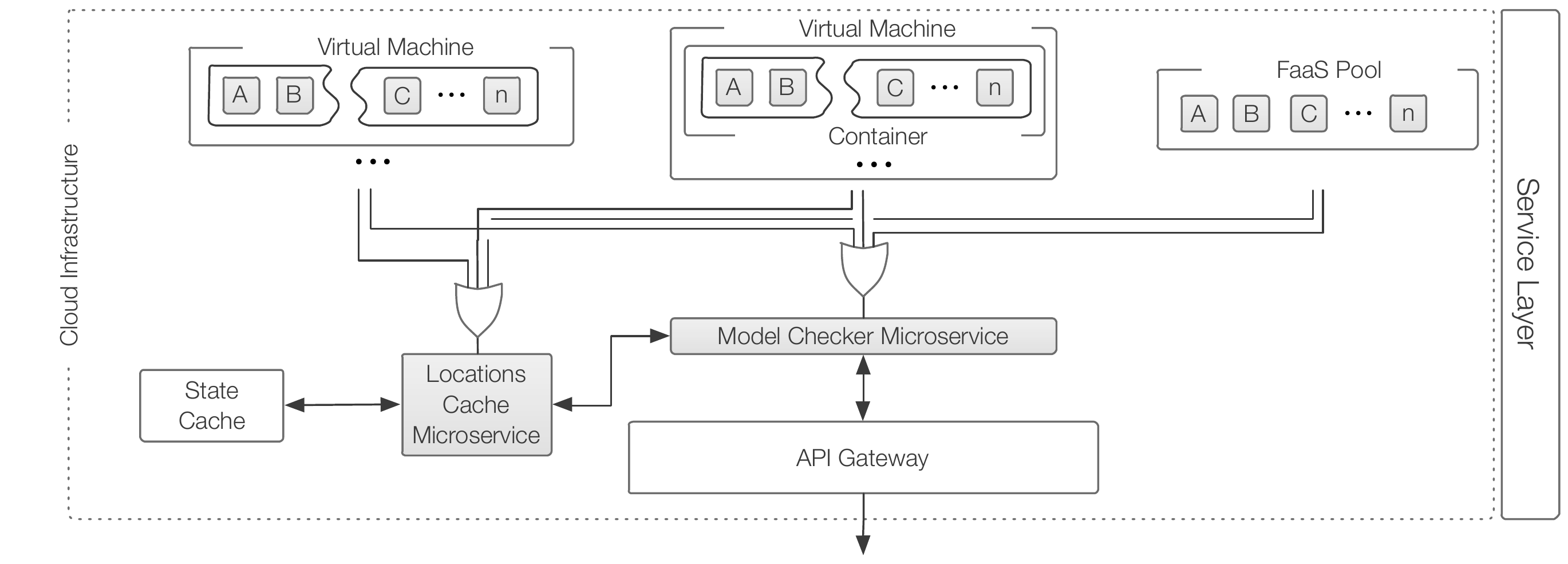}
	\vspace{-.5cm}
\caption{Deployment diagram for the service layer.}
	\label{fig:arch-iot-micro}
\vspace{-.2cm}
\end{figure}


The service-based architecture consists of two basic building blocks, materialized as
two separate microservices\footnote{Complete source code as well as accompanying evaluation models are available at~\cite{paperstuff}.}
Stateless components reflecting solely computation (i.e., the model checking procedure evaluating spatial properties of Section~\ref{sec:spaceverif}) and mostly immutable data (i.e., the topological graph of space of Section~\ref{sec:space-evalmodel}) form the \texttt{model checker} microservice, which receives the current system state and a property, and verifies if such property holds. 
The system state refers to the current spatial distribution of entities in the city as a closure space, while the property captures the QoS requirements that should be evaluated upon the current state of the system.  
 Stateful components form the \texttt{location-cache} microservice, which provides an interface to the current global state of the space, i.e., the location of the devices at a given point in time. Needless to say, both microservices provide HTTP/RESTful interfaces. 
In this manner, those microservices can be reused, scaled and maintained individually; therefore, the software and hardware components in the IoT deployment can be reused and upgraded efficiently. 
In the following, we propose concrete deployments and instantiations of those microservices with the goal of enabling computation offloading to the cloud, and then discuss their advantages and disadvantages. Evaluation of the different deployments upon a real-world workload is provided in Section~\ref{sec:evaluation}.

\subsection{Elementary Cloud Deployments}

\paragraph{Monolithic Deployment}
\label{sec:monolith}

The whole logic of the application (i.e., the two microservices and the database) lives within a single deployment unit. This is the straightforward option for the cloud as it can be deployed in a single monolithic server, typically a virtual machine (VM), which substantially reduces the operational burden~\cite{lewis2015monolith}. However, scalability is constrained in two ways: first, given the limited computing resources of such a monolithic deployment, and second, because the scale unit is the whole application, even though different microservices have unequal loads and may need to scale separately.
Alternatively, the monolith can be deployed in a multi-server environment, with a load balancer to distribute the load among multiple VMs. However, adding VMs to handle workload bursts may take several minutes~\cite{Quatrocchi2016discrete}, which can cause severe SLA violations. Again, the fact that all components scale together (despite their actual utilization) multiplies the resource overprovisioning, something not cost-effective.

\paragraph{Containerized Microservices Deployment}

Each microservice may be developed separately using different technological stacks, deployed and scaled independently in a containerized manner. 
Containers are executed directly on top of the host operating system (typically a VM), optionally with the help of a container manager (e.g., Docker~\cite{docker}). Each container typically hosts one microservice, with its own platform stack and application code. Containers have various advantages when compared to VMs: they are more lightweight, and faster to boot and terminate~\cite{FelterContainerVm15}. This allows for faster and more reactive scalability: if scaling VMs takes several minutes, adding a container or changing its resources takes seconds. Since multiple applications that share the same virtualized resources may see different workloads, the finer granularity and the higher adaptation speed of working with containers allow us to use resources more efficiently~\cite{Quatrocchi2016discrete}. The disadvantages of this approach may rely on the increased operational burden (since there are several deployment units to manage and scale) and, still, SLA violations that may occur during scaling-up actions. 

\paragraph{Functions-as-a-Service Deployment}

FaaS, also known as serverless~\cite{Roberts:2016,MateosFaaster17,Hendrickson:2016}) has been recently proposed as an alternative cloud paradigm in which business functionality is provided without pre-allocating computational resources. Instead, shared resources (e.g. containers) are used to provision and execute functions on demand, typically in a few milliseconds. Therefore, microservices are implemented as functions (often called \emph{lambdas}~\cite{awslamda}) allowing a straightforward deployment to the cloud. 
The horizontal scaling in FaaS is completely automatic, elastic, and more reactive than the typical solutions of scaling a virtual machine or spinning up containers against bursts of workload~\cite{Hendrickson:2016}. A disadvantage of this approach may be vendor lock-in, since specific cloud solutions (e.g. AWS Lambda\cite{awslamda}, IBM/Apache Openwhisk\cite{openwhisk}, Azure Functions\cite{azurefunctions}) are tied to the service ecosystem from the same cloud provider. Additionally, functions are stateless by definition, which limits the applicability of the model -- stateful microservices should store the state outside of the function itself, as done for the \texttt{location-cache} in our deployment. 
Since the computationally-intensive model checking workloads we investigate require minimal state, a FaaS deployment fits such workload requirements particularly well.

\vspace{1cm}
\subsection{Spatial Analysis Workloads over a Hybrid Deployment}

\label{sec:exp-hybrid}

 \begin{wrapfigure}{r}{0.5\textwidth}
	\centering
	\includegraphics[width=0.51\columnwidth]{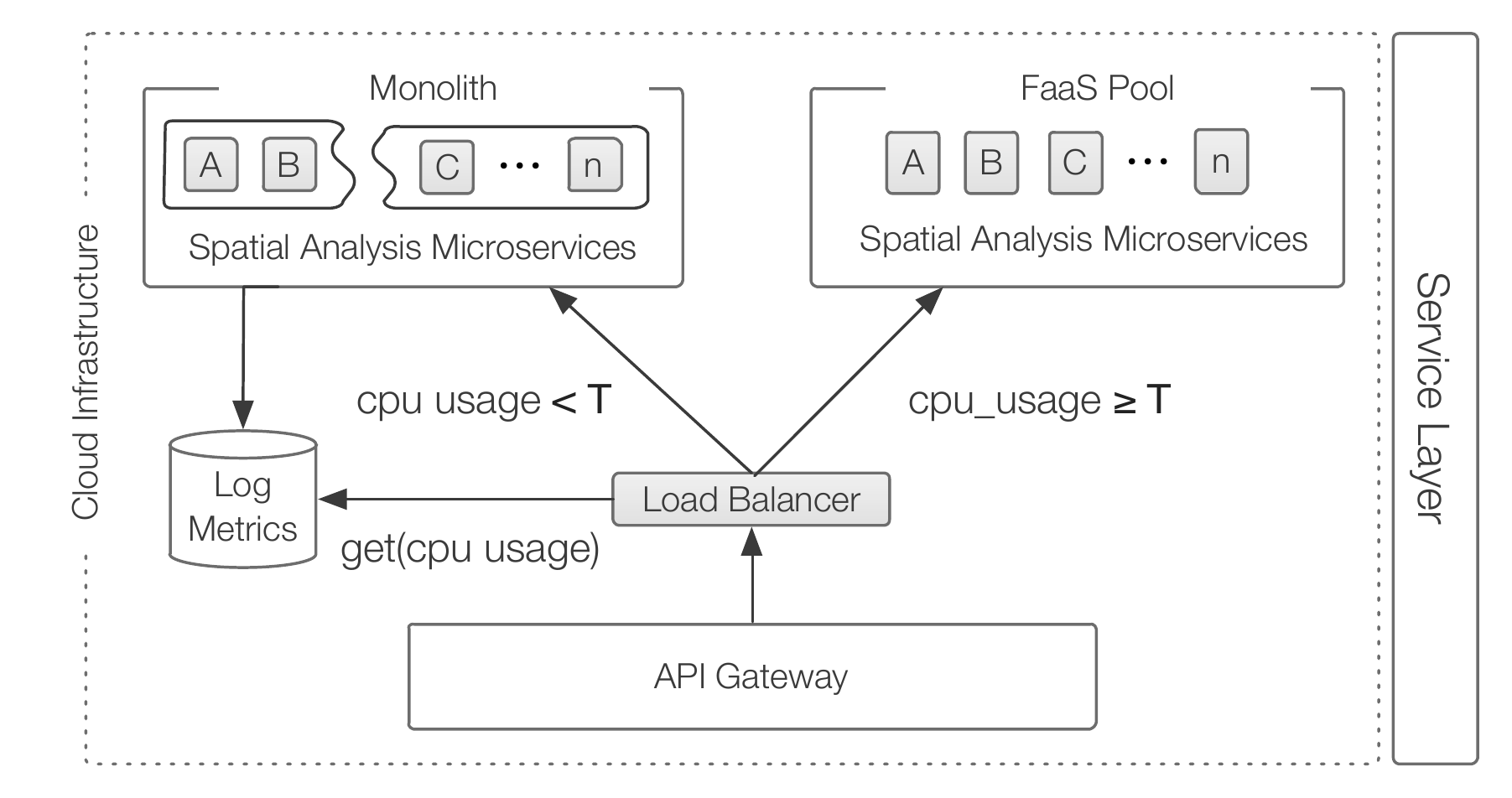}  
	\caption{Sample Hybrid deployment featuring FaaS and a monolith server behind a load balancer.}
	\label{fig:hybrid-arch}
\end{wrapfigure}

A deployment based on VMs or containers, while reducing operational burden may fail to perform with respect to SLAs due to slow scaling actions. On the other hand, the elasticity that FaaS offers should be leveraged. 
To this end, we propose a novel \textit{hybrid} deployment: a baseline infrastructure based on VMs or containers to serve normal, expected workloads, plus a FaaS fallback infrastructure to serve unexpected workload peaks. The rationale is to  leverage lower cost of usual scaling options while fluctuations are managed by highly elastic and reactive FaaS components. Figure~\ref{fig:hybrid-arch} shows our proof-of-concept hybrid deployment featuring a Monolithic server plus a FaaS pool. Instead of invoking directly microservices, a Load Balancer is in charge of forwarding the request to the corresponding server or function, depending whether the CPU load of the server is higher than a given threshold \textit{T}. The request is then forwarded to a lambda function when the server is overloaded and thus  may fail to process the request. In the following section, we evaluate this the alternative deployments with a sample workload scenario.

%% file: sections/evaluation.tex

\section{Evaluation}
\label{sec:evaluation}

For evaluating the proposed approach and deployments, we developed tool support and a proof-of-concept implementation based on the SLCS topochecker~\cite{ciancia2014specifying}, available online along with accompanying material~\cite{paperstuff}. We deployed the prototype to the different cloud alternatives discussed in Section~\ref{sec:deployment}.
Our evaluation goals target realization and applicability of our approach for spatial service-based analysis for IoT applications. Concretely, we aim to:

\begin{itemize}
	\item Investigate feasibility of the approach through concrete deployments in the cloud;
	\item Compare different deployments in terms of latency and SLA in the context of a normal workload, i.e., one known beforehand, to provide a baseline for the required computational resources and cost on each alternative, and
	\item Perform stress tests over the different deployments by increasing the workload rate, to assess scalability of each alternative.
\end{itemize}
 

\noindent
Our evaluation model and workload is sourced from the Microsoft T-Drive trajectory dataset~\cite{yuan2010t,yuan2011driving}, which contains trajectories of 10,357 taxis in Beijing spanning one week. The number of coordinate points and total trajectories distances in the dataset is in the order of millions.
Based on the T-Drive dataset, we derive a closure model as per Section~\ref{sec:space-evalmodel}. 
We source POIs of Beijing from the OpenStreetMap repository~\cite{openstreetmap}.
To obtain discrete trajectory data from the trajectory dataset, we record presences of entities (taxis) over time and space -- i.e., over the points of the accessibility graph previously defined. Specifically, for every single taxi in the dataset, its trajectory -- as a line -- is placed over the POIs of Beijing. If the distance from a point in the trajectory to a POI is less than a predefined length (10 meters for our evaluation purposes), a time-stamped presence of the taxi in that POI is recorded. 
The result is a discrete, timestamped sequence of presences of entities in POIs, which ultimately leads to a graph capturing accessibility of such POIs in geographical space.
For example, if a taxi starts from a POI "Tsinghua University", goes through a bus stop and arrives at "Peking University", then the three are considered (step-wise) connected, since accessibility was demonstrated by at least one trajectory.  The predefined distance aims to mitigate sensing errors in the data, as well as resolution within the scale of which points-of-interest are defined. 
For the T-Drive dataset, an example trace would record $\{taxiA, date , TsinghuaUniversity\} \cdot \{taxiA, date , BusStopA\} \cdot \{taxiA, date , PekingUniversity\}$. 
Evaluation datasets and derived models are available in accompanying material~\cite{paperstuff}.

We present traditional deployments setup on Section~\ref{sec:eval-setup}, and the experimental results obtained in Section~\ref{sec:eval-experiments}. Then, we evaluate the hybrid deployment alternative in Section~\ref{sec:exp-hybrid}, and subsequently discuss threats to validity.

\subsection{Experiments Setup}
\label{sec:eval-setup}

Given a closure model and a trace of presences as previously defined, we proceed to consider the workload induced by the active entities in the space -- in this case the Beijing taxis -- given the scenario presented. We assume that each registered presence of a taxi in a POI, also implies the evaluation of certain spatial properties (similarly to Section~\ref{sec:motivation-example}) by calling the \texttt{model checker} microservice. The evaluation of the property is based on the global state of \emph{all} taxis in the Beijing dataset at that time.

 \begin{wrapfigure}{r}{0.5\textwidth}
 	\centering
 	\includegraphics[width=0.5\textwidth]{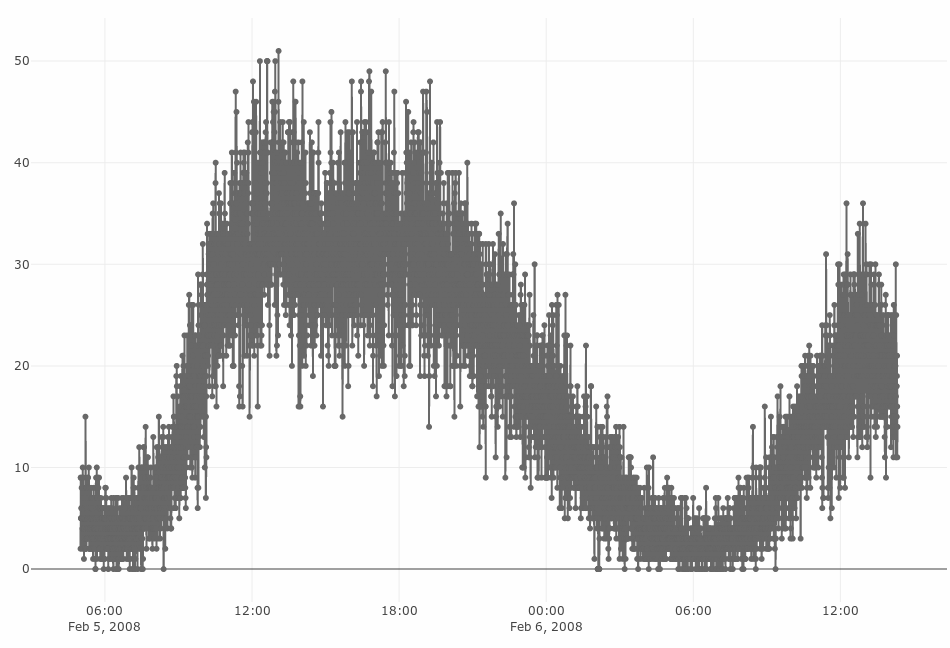}  
 	\caption{Beijing traces~\cite{yuan2010t} dataset evaluation workload~\cite{paperstuff} as taxi presences (y-axis) near points-of-interest~\cite{openstreetmap} over the course of hours of the days of Feb 5-6 2008 (x-axis).}
 	\label{fig:sine}
 \end{wrapfigure}

We limit our evaluation clock-time to a window of 24 hours in Beijing. Figure~\ref{fig:sine} depicts single requests (i.e. taxi presence heartbeats) over time for a representative weekday, where one can observe that the workload exhibits a sine wave following the working hours.
We consider a closure model of 36805 edges, with 15456 propositions over 5152 nodes~\cite{paperstuff}. 

The performed experiments were threefold. We defined an SLA of 30 seconds for the response time of the properties evaluation service. For the first experiment, we executed the sample window of 24 hours of Figure~\ref{fig:sine} and profiled the necessary infrastructure (for each deployment) in order to satisfy the SLA within the whole period. This allowed us to determine the baseline infrastructure and cost to serve a typical workload based on the evaluation model.
Then, in the second experiment we proceeded with a stress test, by reproducing one hour of workload with different time multipliers, i.e., simulating that the events arrive with a faster rate. This allowed us to assess the scalability of each deployment, in a scenario in which the workload rate is  increased with time multipliers. 
Finally, the third and last experiment assesses the performance of our hybrid approach with respect to using only VMs or FaaS. The hybrid combines the best of the previous alternatives: a baseline infrastructure with preallocated resources (in the form of VMs or Containers) to serve normal workloads, plus a FaaS fallback infrastructure to serve unexpected peaks in the workload.


The example properties to be checked (Formulae~\ref{p1}-\ref{p3}) are system-wide goals in line to the scenario of Section~\ref{sec:motivation-example}, that aim to assign penalties or rewards to taxis in order to avoid clamping in the city.
Specifically, P1 encodes the requirement that a taxi shall be able to reach department stores traversing points of subway or bus stops. However, traversal may not go through a hospital.  P2 encodes that the taxi is always in the vicinity of a hospital or a hotel (by two steps away), but not immediately next to restaurants.
P3 specifies that a taxi should be within an area characterized by proximity to tourist attractions (e.g. the Forbidden Palace) and subways, up to zoo's (e.g. the Beijing Zoo). Such a property, utilizing the $\mathcal{T}$ operator of Section~\ref{sec:spaceverif} could be used to reward taxis implementing a temporary city-wide policy of attracting tourists which are located in the center of the city to go to the zoo, since they will be able to find more taxis within this area. Note that specification formulae utilize the complex operators previously defined -- the corresponding elementary formulae are sizable. 
The propositions used in the specifications are found in the generated dataset available in accompanying material~\cite{paperstuff}.

\begin{small}
      \begin{empheq}[]{align}
      & \textbf{(P1)} \hspace{0.01cm} \mathtt{taxi}\ \mathcal{\Re}\Big(\mathtt{(!HEALTHHOSPITAL\vee \big( \big(TRANSPORTSUBWAY\vee TRANSPORTBUSSTOP\big) \wedge taxi \big) }\Big)\  \mathtt{DEPARTMENTSTORE}. \label{p1} \\
      & \textbf{(P2)} \hspace{0.01cm} \mathtt{taxi} \wedge \mathcal{N}_2 \big(\mathtt{ACCOMMOHOTEL} \vee \mathtt{HEALTHHOSPITAL}\big) \wedge \mathcal{N} \big( \mathtt{FOODRESTAURANT} \big). \label{p2} \\
      & \textbf{(P3)} \hspace{0.01cm} \mathtt{taxi} \wedge \mathcal{N}_2 \big( \mathtt{TOURISTATTRACTION} \wedge \mathtt{TRANSPORTSUBWAY} \big)\ \mathcal{T}\  \mathtt{TOURISTZOO} . \label{p3} 
    \end{empheq} 
\end{small}


We deployed the microservice architecture described in Section~\ref{sec:iot-arch} (Figure~\ref{fig:arch-iot-micro}) in four different environments. We note that the workload, being a model checking procedure, is heavily computation-intensive, but its memory requirements are minimal, so we ignore RAM specifications of the infrastructures for clarity. 

\begin{enumerate}[wide=0pt, leftmargin=2em]
	\item [\textbf{Device:}] Local execution of the model checking load in an IoT resource-constrained device (e.g. a mobile phone), featuring am ARMv6 A53 1GHz CPU.
	\item[\textbf{Monolith:}] Large VM instance containing the two microservices in a single deployment package. The VM ranges from 16 CPUs 
	(c5.4xlarge AWS instance) for light workloads, up to 64 CPUs 
	(c5.18xlarge) for heavy workloads.
	\item [\textbf{Containers:}] AWS Elastic Container Services (ECS\footnote{https://aws.amazon.com/ecs/}), where each microservice is containerized, deployed and scaled independently. The infrastructure ranges from 10 to 25 instances with 2 CPUs 
	each (t2.medium), depending on the workload.
	\item[\textbf{Lambdas:}] Functions-as-a-Service as AWS Lambda functions were configured with 
	~2 vCPUs each\footnote{CPUs are assigned proportionally to the RAM (2GB in this case) -- https://aws.amazon.com/lambda/},
	scaling automatically from zero to the level of parallelism required, according to the workload.
\end{enumerate}

The services deployed correspond to a \texttt{location-cache} microservice as an in-memory key-value store, while
the \texttt{checker} microservice as an encapsulation of the model checker~\cite{ciancia2015experimental}, which receives the current state and a property, and verifies if the property holds.
The IoT device, featuring similar processing power to a mobile phone, encapsulates a model checker procedure compiled accordingly per its software architecture stack. 
The relevant implementations are available in~\cite{paperstuff}.


\subsection{Cost and Scalability Evaluation Results}
\label{sec:eval-experiments}


Table~\ref{tab:results-exp-1} shows the infrastructure and cost results for the first experiment, considering 24 hours of execution (resulting in 8864 model checker invocations) sliced in periods of three hours. The third column (\emph{Cores}) shows the approximate number of cores needed to serve such a workload within the SLA of 30 seconds per call. Then, each deployment is scaled to be close to that number during each period. 

Results show that, when the workload is known beforehand, the Containerized Microservices solution is the most cost effective: \$456.96/month, considering a month as 21.73 working days (the workload considered decreases significantly during weekends). 
Subsequently, the FaaS deployment costs \$771.67/month, calculated upon the amount of memory assigned to the function, and the exact number of invocations and execution time (12 seconds on average), which eases its fine-tuning. 
The less cost effective solution is the Monolithic Server deployment, at a cost of \$932.63/month. This is due to the fact that the Monolithic Server incurs resource over-provisioning most of the time, even though we are adjusting the size of the VM for each period.
Finally, we also profiled the execution in the mobile device, which represents a single taxi in the dataset. Rather than testing with the workload (multiple taxis), we performed sequential calls. 
This captures the case where instead of a cloud solution, one chooses a deployment where properties are evaluated on the devices making up the IoT system. 
Execution results shown that single invocations take 99 seconds on average to complete, which violates our SLA of 30 seconds, making it unfeasible in practice (barring e.g. battery drain effects). Nevertheless, we additionally included the mobile device in the subsequent experiment for completeness.

\begin{table*}
	
	\caption{Baseline infrastructure and cost for the cloud deployments considering
		24h workload.}
	\label{tab:results-exp-1}
	
	{\footnotesize{}}%
	\begin{tabular*}{1\textwidth}{@{\extracolsep{\fill}}>{\raggedright}p{0.8cm}>{\raggedright}p{0.8cm}>{\raggedright}p{0.8cm}|>{\centering}p{1.2cm}>{\centering}p{1.2cm}|>{\centering}p{1.2cm}>{\centering}p{0.8cm}>{\centering}p{0.8cm}>{\centering}p{0.8cm}|>{\centering}p{0.8cm}>{\centering}p{0.8cm}}
		\hline 
		\multirow{2}{0.8cm}{\textbf{\footnotesize{}Hour}} & \multirow{2}{0.8cm}{\textbf{\footnotesize{}Calls}} & \multirow{2}{0.8cm}{\textbf{\footnotesize{}Cores (aprox)}} & \multicolumn{2}{>{\centering}p{2.4cm}|}{\textbf{\footnotesize{}Containerized}{\footnotesize{} }\textbf{\footnotesize{}Microservices}} & \multicolumn{4}{>{\centering}p{4.2cm}|}{\textbf{\footnotesize{}Monolithic}{\footnotesize \par}
			
			\textbf{\footnotesize{}Server}} & \multicolumn{2}{>{\centering}p{2cm}}{\textbf{\footnotesize{}FaaS Deployment}}\tabularnewline
		&  &  & \textbf{\footnotesize{}VMs{*}} & \textbf{\footnotesize{}cost} & \textbf{\footnotesize{}VM} & \textbf{\footnotesize{}cores} & \textbf{\footnotesize{}memory} & \textbf{\footnotesize{}cost} & \textbf{\footnotesize{}sec/call} & \textbf{\footnotesize{}cost}\tabularnewline
		\hline 
		{\footnotesize{}6\textendash 9} & {\footnotesize{}312} & {\footnotesize{}20} & {\footnotesize{}10} & {\footnotesize{}1.4} & {\footnotesize{}c5.4xl} & {\footnotesize{}16} & {\footnotesize{}32} & {\footnotesize{}2.04} & {\footnotesize{}12} & {\footnotesize{}1.25}\tabularnewline
		{\footnotesize{}9\textendash 12} & {\footnotesize{}1283} & {\footnotesize{}40} & {\footnotesize{}20} & {\footnotesize{}2.8} & {\footnotesize{}c5.9xl} & {\footnotesize{}32} & {\footnotesize{}64} & {\footnotesize{}4.6} & {\footnotesize{}12} & {\footnotesize{}5.13}\tabularnewline
		{\footnotesize{}12\textendash 15} & {\footnotesize{}1663} & {\footnotesize{}50} & {\footnotesize{}25} & {\footnotesize{}3.5} & {\footnotesize{}c5.18xl} & {\footnotesize{}64} & {\footnotesize{}128} & {\footnotesize{}9.18} & {\footnotesize{}12} & {\footnotesize{}6.65}\tabularnewline
		{\footnotesize{}15\textendash 18} & {\footnotesize{}1676} & {\footnotesize{}50} & {\footnotesize{}25} & {\footnotesize{}3.5} & {\footnotesize{}c5.18xl} & {\footnotesize{}64} & {\footnotesize{}128} & {\footnotesize{}9.18} & {\footnotesize{}12} & {\footnotesize{}6.70}\tabularnewline
		{\footnotesize{}18\textendash 21} & {\footnotesize{}1668} & {\footnotesize{}50} & {\footnotesize{}25} & {\footnotesize{}3.5} & {\footnotesize{}c5.18xl} & {\footnotesize{}64} & {\footnotesize{}128} & {\footnotesize{}9.18} & {\footnotesize{}12} & {\footnotesize{}6.67}\tabularnewline
		{\footnotesize{}21\textendash 00} & {\footnotesize{}1202} & {\footnotesize{}40} & {\footnotesize{}20} & {\footnotesize{}2.8} & {\footnotesize{}c5.9xl} & {\footnotesize{}32} & {\footnotesize{}64} & {\footnotesize{}4.6} & {\footnotesize{}12} & {\footnotesize{}4.80}\tabularnewline
		{\footnotesize{}00\textendash 03} & {\footnotesize{}705} & {\footnotesize{}30} & {\footnotesize{}15} & {\footnotesize{}2.1} & {\footnotesize{}c5.4xl} & {\footnotesize{}16} & {\footnotesize{}32} & {\footnotesize{}2.04} & {\footnotesize{}12} & {\footnotesize{}2.82}\tabularnewline
		{\footnotesize{}03\textendash 06} & {\footnotesize{}355} & {\footnotesize{}20} & {\footnotesize{}10} & {\footnotesize{}1.4} & {\footnotesize{}c5.4xl} & {\footnotesize{}16} & {\footnotesize{}13} & {\footnotesize{}2.04} & {\footnotesize{}12} & {\footnotesize{}1.42}\tabularnewline
		\hline 
		{\footnotesize{}Day} & {\footnotesize{}8864} & \multicolumn{1}{>{\raggedright}p{0.8cm}}{} &  & \multicolumn{1}{>{\centering}p{1.1cm}}{{\footnotesize{}21}} &  &  &  & \multicolumn{1}{>{\centering}p{0.8cm}}{{\footnotesize{}42.86}} &  & {\footnotesize{}35.46}\tabularnewline
		{\footnotesize{}Month{*}{*} } & {\footnotesize{}192614} & \multicolumn{1}{>{\raggedright}p{0.8cm}}{} &  & \multicolumn{1}{>{\centering}p{1.1cm}}{{\footnotesize{}456.96}} &  &  &  & \multicolumn{1}{>{\centering}p{0.8cm}}{{\footnotesize{}932.63}} &  & {\footnotesize{}771.67}\tabularnewline
		\hline 
		\multicolumn{11}{l}{{\footnotesize{}{*}t2.medium AWS instance, 2 CPUs, 4GB memory}}\tabularnewline
		\multicolumn{11}{l}{{\footnotesize{}{*}{*}Month = 21.73 working days}}\tabularnewline
	\end{tabular*}{\footnotesize \par}
\end{table*}


For the second experiment we considered a representative lapse of one hour of workload, from 11am to 12am of February 5 (Figure~\ref{fig:sine}), which is the beginning of a daily peak comprising 536 data points. We executed stress tests over the different Cloud deployments, by multiplying the rate in which requests arrive (one per each data point). Each request triggers one invocation of the model checking procedure. We stretched the 536 invocations in 12 minutes (5x faster rate), 6 minutes (10x), 3 minutes (20x), 1.5 minutes (40x) and 1 minute (60x). Then, we measured the total time per invocation, considering the SLA of 30 seconds defined before, and disaggregated it by computation time and wait time, i.e., when the request is queued waiting for computation resources. 

\begin{figure*}[htb]

\begin{subfigure}{5cm}
	\centering
	\includegraphics[width=.9\columnwidth]{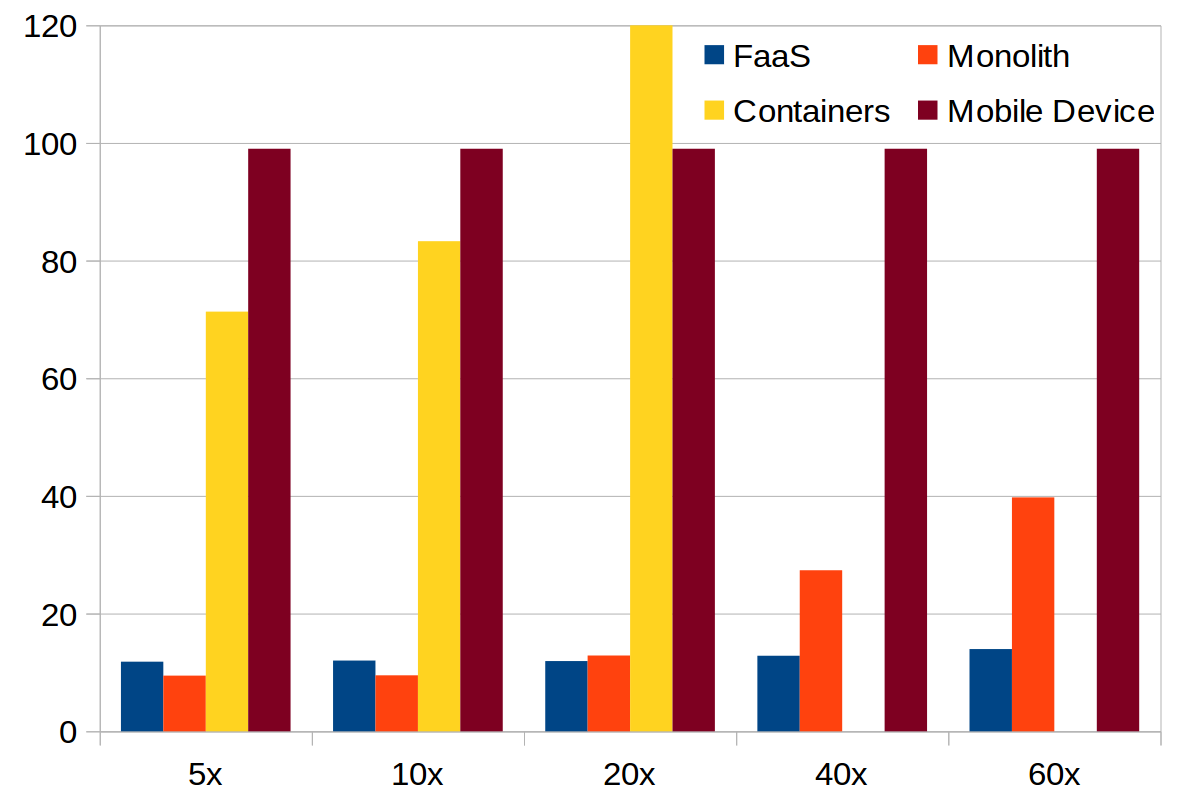} 
	\vspace{-0.2cm}
	\caption{Total time per request (a)}
	\label{fig:total-time}
\end{subfigure}
~
\begin{subfigure}{5cm}
	\centering
	\includegraphics[width=.9\columnwidth]{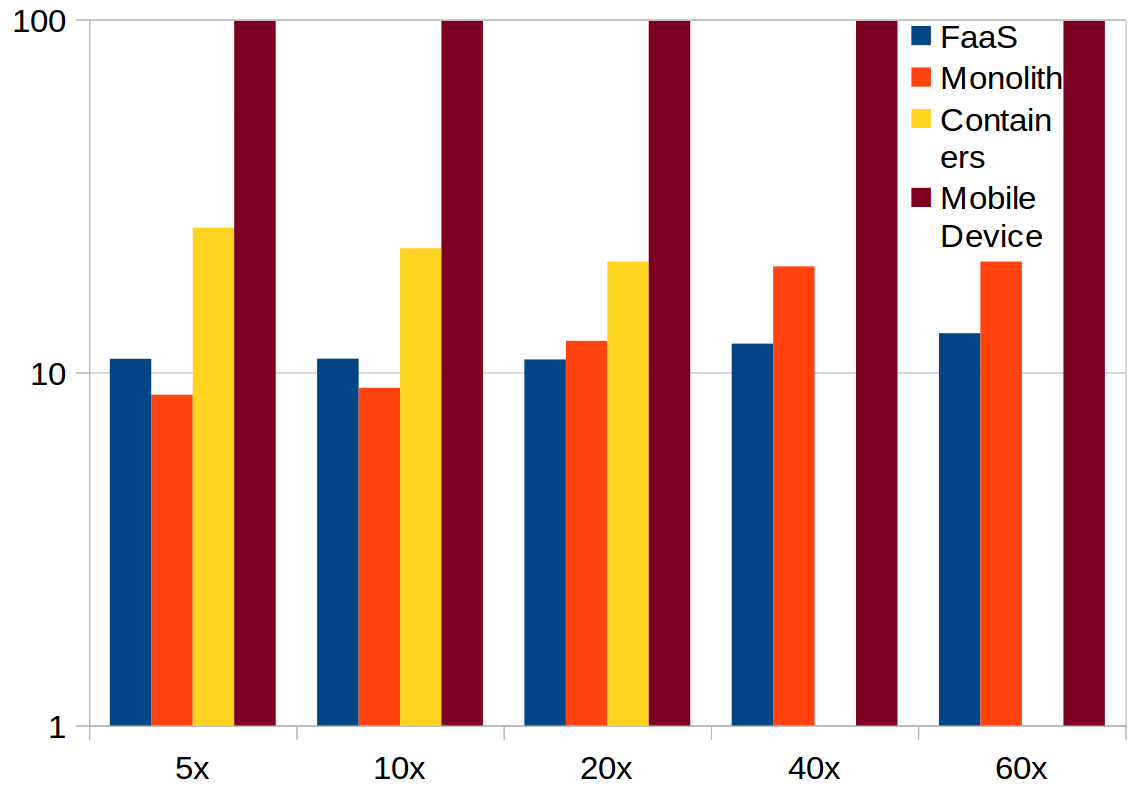} 
	\vspace{-0.2cm}
	\caption{Computation time (logarithmic scale)}
	\label{fig:computation-time}
\end{subfigure}
~
\begin{subfigure}{5cm}
	\centering
	\includegraphics[width=.9\columnwidth]{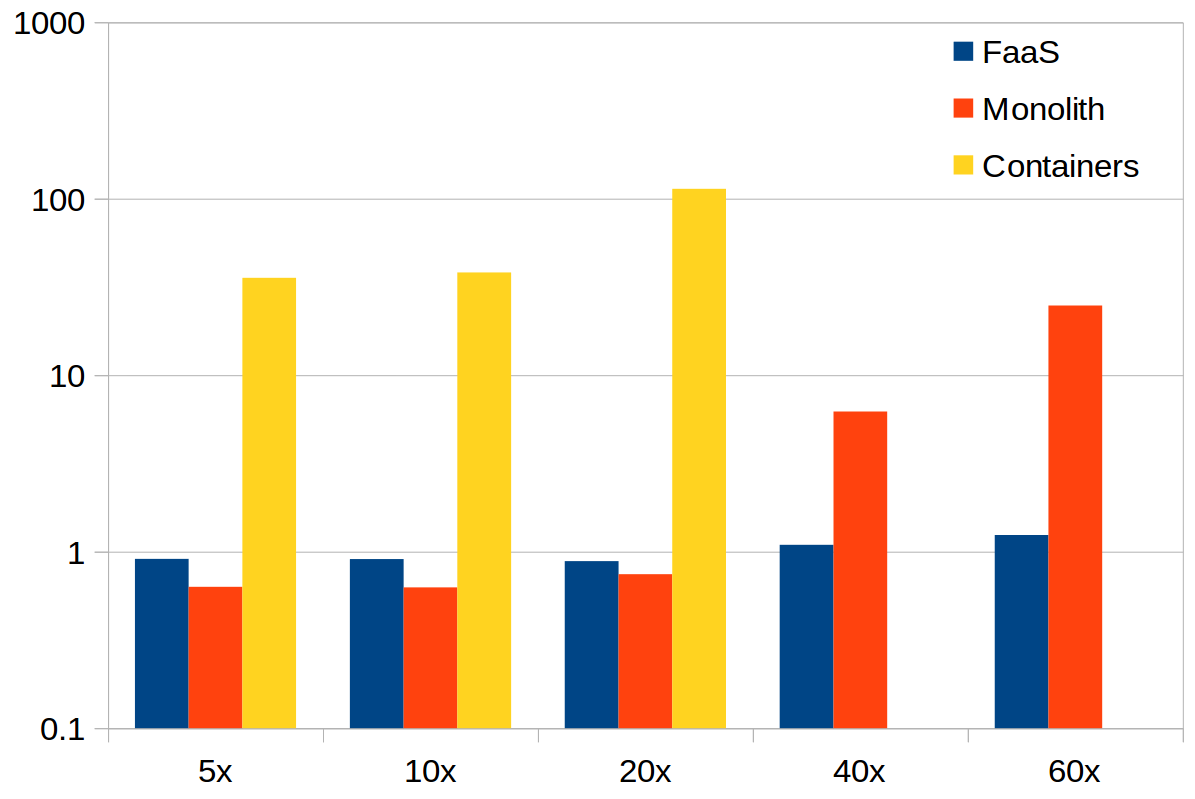} 
	\vspace{-0.2cm}
	\caption{Wait time (logarithmic scale).}
	\label{fig:wait-time}
\end{subfigure}
\vspace{-0.5cm}
\caption{Average time per request for different deployments and workflow rates.}
  \vspace{-.4cm}
\end{figure*}

Figure~\ref{fig:total-time} depicts the median time per invocation for the different rates and deployments. Additionally, we plotted the mobile device (rightmost column), which executes sequential requests independently of the rate: as soon as one request is processed, another one is fired. 
The Containerized Microservices in the ECS cluster presented the worst performance overall, with requests timing out (error 504) with 20x rate or more, which explains the absence of the yellow column for 40x and 60x rates.  Since a scaling-up action (that can demand several seconds~\cite{Quatrocchi2016discrete}) triggers when more containers are needed to handle the workload, requests timeout in the meantime.
The FaaS alternative held an (almost) constant performance with 12 seconds per invocation for all workloads. Even though commercial FaaS solutions can give the impression of "infinite" scalability, the tradeoff between resource availability and cost (Table~\ref{tab:results-exp-1}) has to be carefully considered.
The Monolithic deployment performed better (between 9 and 12 seconds per invocation) up to 20x rate, then started to decrease its performance due to resource unavailability. As discussed in Section~\ref{sec:monolith}, the monolith could also scale in a multi-server environment, although adding VMs to handle workload bursts as in this experiment may take several minutes~\cite{Quatrocchi2016discrete}, which can cause severe SLA violations in the meantime.
Finally, the Mobile Device computes the requests sequentially with a constant performance of an average of 99 seconds per invocation. Needless to say, such processing time is prohibitive as it violates the defined SLA. 
Figures~\ref{fig:computation-time} and \ref{fig:wait-time} show the actual computation and waiting time respectively. Recall that waiting occurs when no computation resources are available at the moment the service is invoked. One can observe that for all the deployments, computation time remains almost constant, the underlying issue being the wait time which substantially increases for faster workload rates. In such cases, requests are queued waiting for available computational resources, which decreases the overall system performance.




From the two first experiments we can highlight the following. First, the solution based on the mobile device, even though it processes only its own requests, has a prohibitive computation time of 99 seconds on average, making it unsuitable in practice. The unfeasibility of performing computation locally calls for computation offloading to the cloud, with all its different flavors. 
The two solutions that performed better were FaaS and Monolithic. The FaaS deployment using AWS Lambda held a constant performance even under dense workloads, being more reactive and scalable than the Containerized and Monolithic alternatives. 
The option of the Monolithic deployment in traditional AWS VMs, performed well for medium workloads, but not against peaks or faster rates, specially when the latter are unknown beforehand.
Finally, the Containerized Microservices deployment, materialized through an AWS ECS Cluster did not seem suitable for this scenario, mainly because of scaling-up delays (even when containers take only seconds to bootstrap).
Regarding the overall cost, when the workload is more or less known (e.g., a normal workday in our taxis scenario) FaaS become more costly in comparison to other deployments that allow one to preallocate resources in advance. This opens the challenge to consider the hybrid approach, as follows.

\textbf{Hybrid vs. monolithic and FaaS deployment.}
For the third experiment, we implemented the hybrid deployment of Section~\ref{sec:exp-hybrid} (Figure~\ref{fig:hybrid-arch}) and assessed its performance with respect to other cloud deployments -- namely monolith and FaaS. Our proof-of-concept hybrid deployment features a Monolithic server plus a FaaS pool, transparent for the end user thanks to the API Gateway, which provides a single REST endpoint to reach any of the checker microservices. Instead of calling directly the microservice, the API Gateway fires a lambda function which acts as the Load Balancer that forwards the request to the corresponding server/function, depending on the load of the server being higher than the configurable threshold \textit{T}.

\begin{figure*}[htb]
	
	\begin{subfigure}{5cm}
		\centering
		\includegraphics[width=.9\columnwidth]{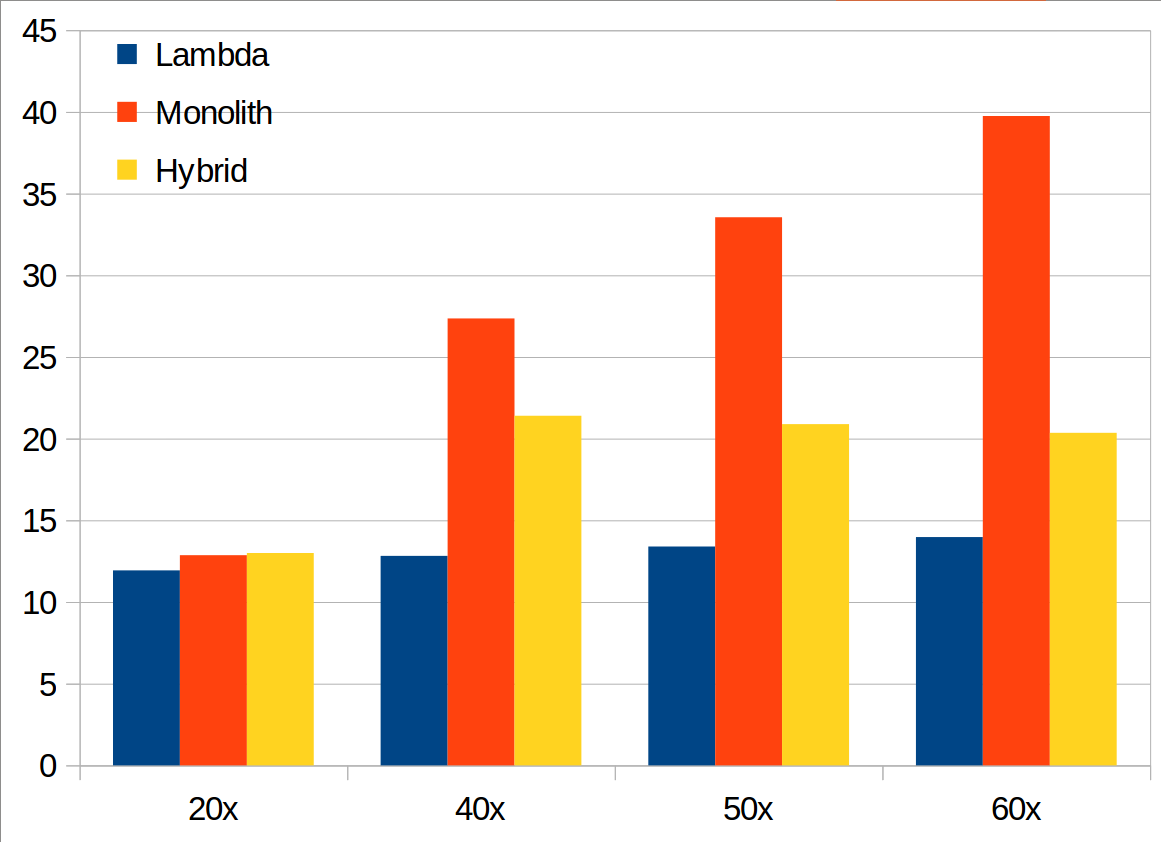} 
		\vspace{-0.2cm}
		\caption{Total time per call (seconds)}
		\label{fig:total-time-exp3}
	\end{subfigure}
	~
	\begin{subfigure}{5cm}
		\centering
		\includegraphics[width=.9\columnwidth]{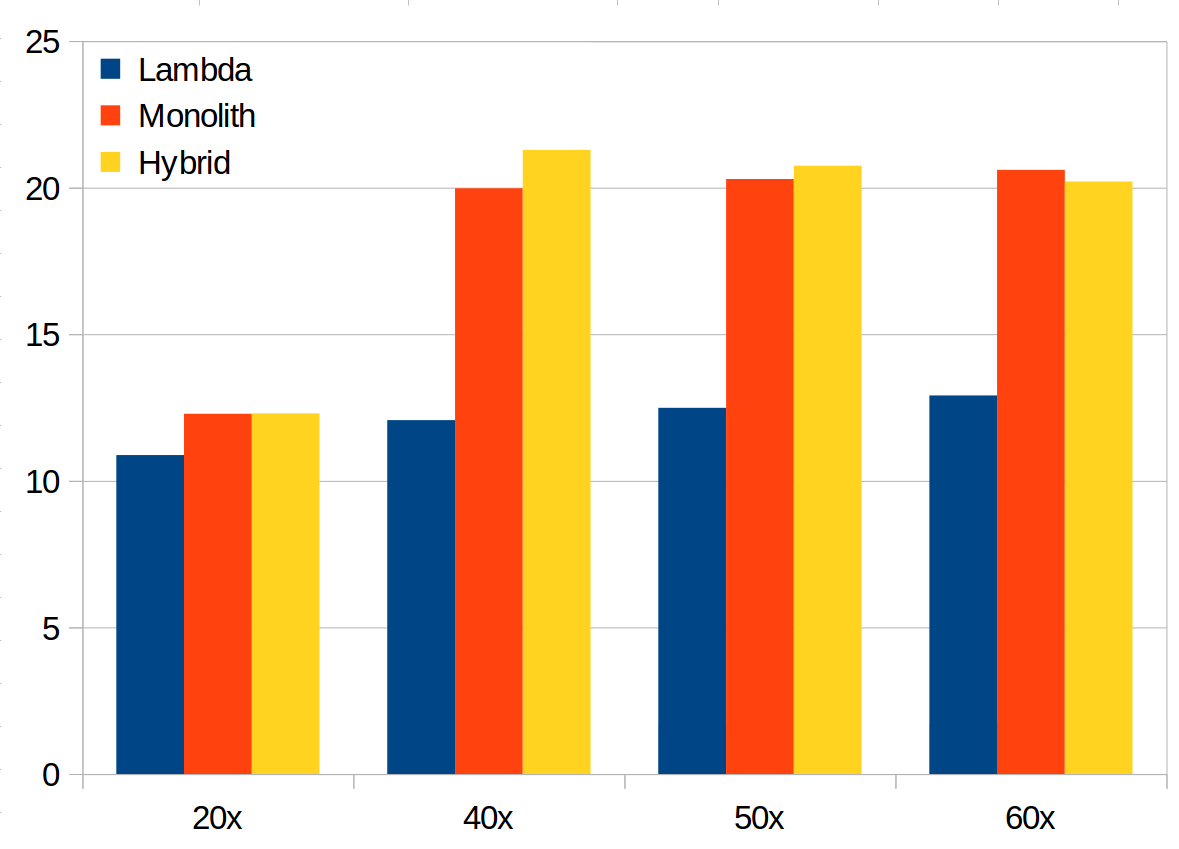} 
		\vspace{-0.2cm}
		\caption{Computation time per call (seconds)}
		\label{fig:computation-time-exp3}
	\end{subfigure}
	~
	\begin{subfigure}{5cm}
		\centering
		\includegraphics[width=.9\columnwidth]{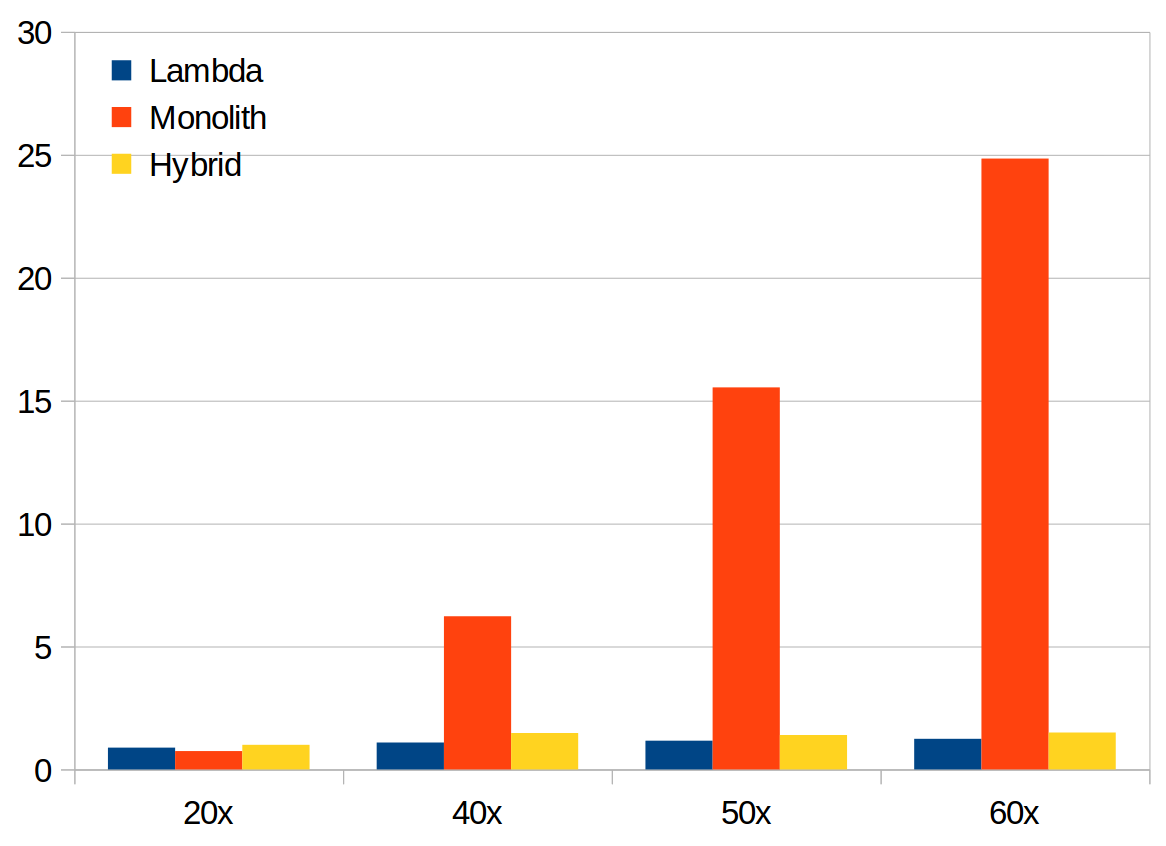} 
		\vspace{-0.2cm}
		\caption{Wait time per call (seconds)}
		\label{fig:wait-time-exp3}
	\end{subfigure}
	\vspace{-0.3cm}
	\caption{Time (y-axis) comparison for different deployments and workflow rates (x-axis).}
	\label{fig:exp3}
  \vspace{-.4cm}
\end{figure*}

Figure~\ref{fig:exp3} depicts the total, computation and wait time per invocation for the different request rates, considering the hybrid deployment, plus lambda and monolith as the two best alternatives according to the previous experiments (see Section~\ref{sec:eval-experiments}). Note that the plotted values correspond to 20x rates (3 requests/second) and higher, since for lighter workloads the hybrid deployment is not necessary (100\% requests being forwarded to the monolithic server). For 40x rates, the monolith served 72\% of requests and FaaS 28\%. For 50x rates, 67\% and 33\% respectively, and for 60x the distribution was 65\% and 35\%. 
For 20x rates all deployments performed equally. For faster rates, the monolith server exponentially decreases its performance mainly due to wait times (insufficient computational resources). Meanwhile, lambda and hybrid held an almost constant performance of 12 and 20 seconds per call respectively.

\begin{figure*}[htb]
	\centering
	\begin{subfigure}{5cm}
		\centering
		\includegraphics[width=1\columnwidth]{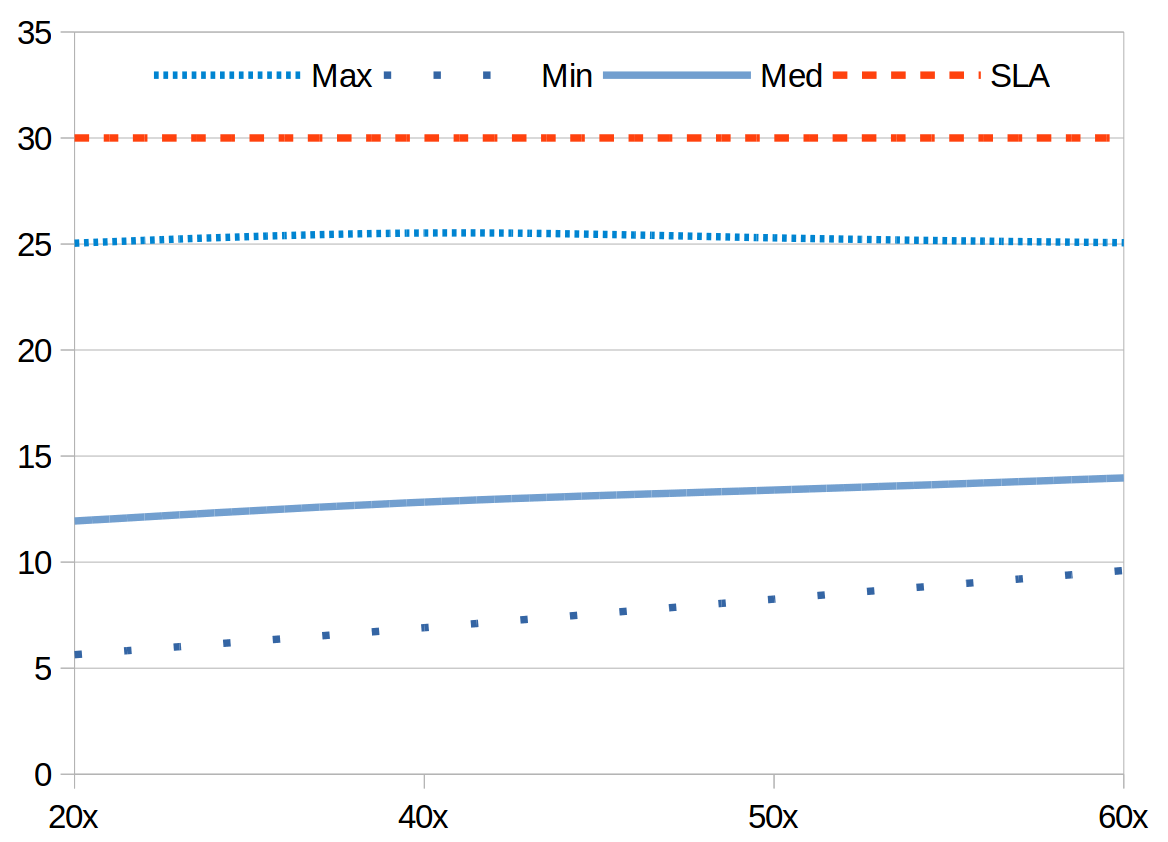}
		\caption{FaaS Deployment.}
		\label{fig:lambda-time}
	\end{subfigure}
	~
	\begin{subfigure}{5cm}
		\centering
		\includegraphics[width=1\columnwidth]{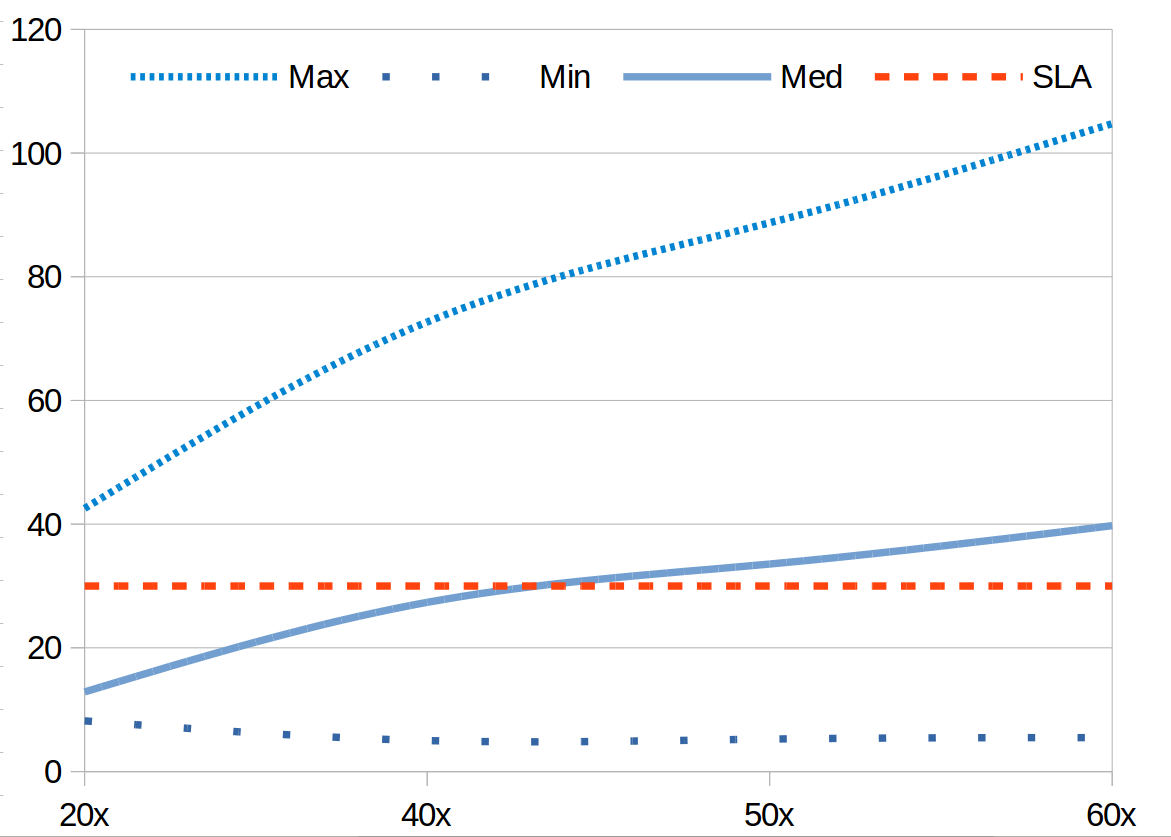}
		\caption{Monolithic Deployment.}
	\end{subfigure}
	~
	\begin{subfigure}{5cm}
		\centering
		\includegraphics[width=1\columnwidth]{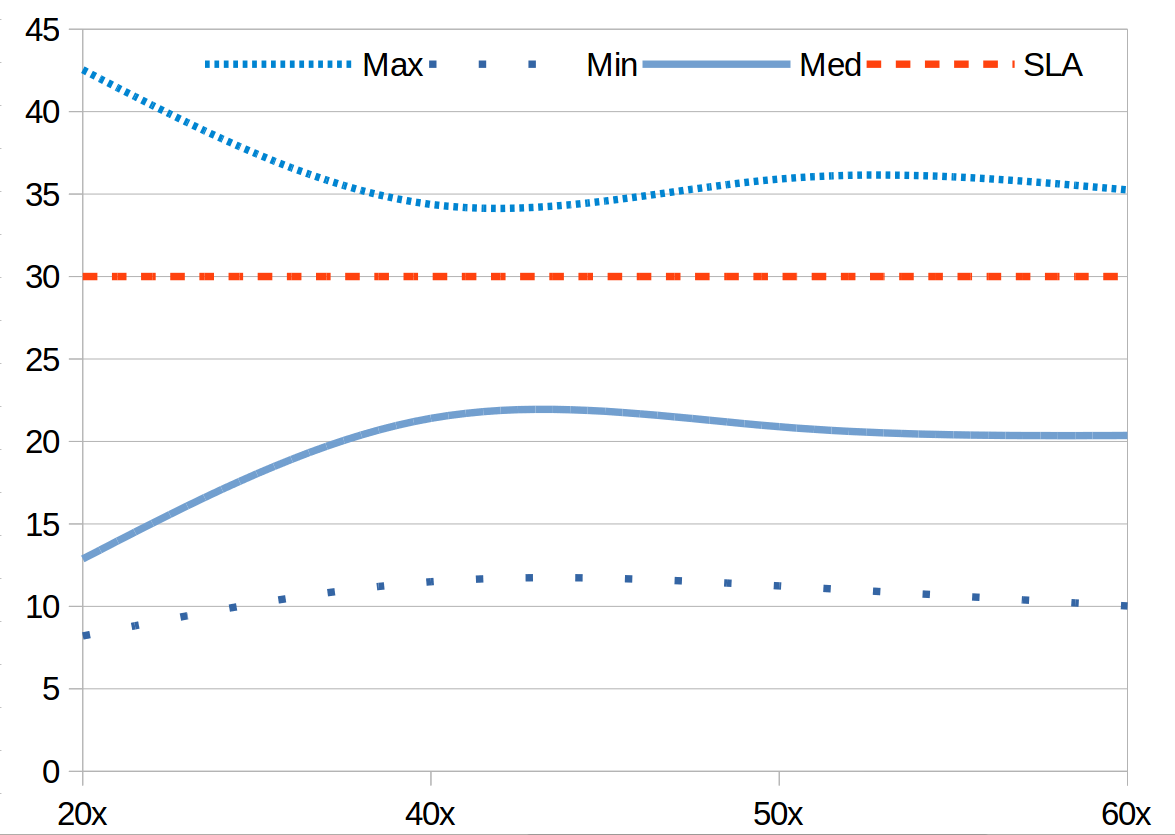}
		\caption{Hybrid Deployment.}
	\end{subfigure}
	\vspace{-0.4cm}
	\caption{Total times (max, min, median) and SLA violations for FaaS, Monolithic and hybrid deployments.}
	\label{fig:lambda-monolith-time}
  \vspace{-.4cm}
\end{figure*}

Figure~\ref{fig:lambda-monolith-time} shows the maximum, minimum and median times per call for FaaS, Monolith and Hybrid deployment, regarding the 30 seconds SLA previously defined. FaaS holds a constant performance without incurring SLA violations. The Monolithic alternative incurs SLA violations for 40x rates or higher, not only for the maximum but also for the median times. Meanwhile, the hybrid deployment shows SLA violations only for the maximum times, but not for the median. This is due to the requests that arrive at the beginning of a peak, that is, when the server starts to be overloaded, but the LB has not yet detected it. 
Conclusively, the hybrid deployment combines the best of the two worlds: preallocated resources to serve normal workloads, while lambda functions allow to serve unexpected peaks on the workload in a reactive and cost-effective fashion. In all cases, the lambda load balancer introduces little overload, less than 300ms per request on average.

Finally, in terms of threats to validity is it worth mentioning that the experiments targeted only one sample model checker~\cite{ciancia2014specifying}, and three spatial properties. Further tests may be needed, considering other properties and implementations that may affect processing times. 
Additionally, one may argue that the defined SLA of 30 seconds is arbitrary. This should be considered only as a reference value in the context of our experiments, whilst an actual SLA may be even lower. In those cases, the model checker implementation and deployment configuration (e.g., timeouts, location cache lifetime, data locality) should be optimized to meet the requirements of a real-time application.
Regarding the workload behavior, the cloud deployment alternatives that perform resource preallocation (i.e., monolith server, containers) require to know the workload beforehand. When this is not possible, it may be better to rely on more flexible deployments such as FaaS and hybrid (the latter with a minimal amount of preallocated resources), which can prove more expensive in the long term. It should be also taken into account that the hybrid deployment used for the third experiment constitutes a proof-of-concept. Fine-tunning the deployment can result in further improvements, in particular regarding the threshold value \textit{T} for CPU usage, and caching policies for the load balancer. Last but not least, using containers and/or edge servers for deploying the hybrid alternative could substantially improve its performance~\cite{Quatrocchi2016discrete,mendonca18toit,GarrigaMendonca2017}, these options being considered as a significant avenue of future work.
 

%

%% file: sections/relatedwork.tex

\section{Related Work}
\label{sec:relatedwork}

We presented an approach for model checking requirements predicating on the spatial distribution of devices comprising an internet-of-things
system at runtime, leveraging a cloud architecture based on microservices.
Consequently, we classify related work into
three categories. 
First, we review foundational work on spatial reasoning, positioning our work. Then, we discuss relevant approaches that have considered model checking and computational intensive tasks using cloud infrastructures. Later, we discuss related engineering approaches that aim at systems utilizing spatial verification for various purposes, in context of which our technical framework may be used.

Topological relations have been traditionally considered in the context of database systems, 
query languages~\cite{kennedy1996survey} and logics for spatial data 
analysis in geographical information systems~\cite{bivand2013spatial}, focusing on the elements of a geometric model~\cite{egenhofer1990categorizing,egenhofer1989topological}. In addition, spatial logics have also been studied in the context of process calculi~\cite{cardelli2002spatial}, where the typical theme is predication against the structure of agents, also with applications to graph databases~\cite{cardelli2001spatial}.
Our choice of a spatial logic over closure spaces~\cite{ciancia2015experimental} as a 
basis for our spatial reasoning support is that closure spaces are a generic mathematical concept serving as the 
interface between arbitrary binary relations. We further note that quantitative reasoning in our application domain appears promising~\cite{bortolussi2014specifying}, something we identify as future work.

The recent extensive availability of cloud processing platforms has fueled notable approaches for classical model checking~\cite{lerda1999distributed,bellettini2013distributed}, proposing a technological transition to exploit the new available architectures. Approaches target classical, foundational verification of temporal logics such as CTL and LTL~\cite{clarke1999model}. The prevalence and availability of the cloud has created an increasing interest in parallelizing and distributing verification techniques. 
Generally, the explotation of cloud resources aims at 
increasing the memory available and reducing the overall time
required by verification by employing distributed techniques~\cite{evangelista2011parallel,barnat2003parallel} or by splitting the given state space into several partial state spaces~\cite{brim2005assumption}. 
We note that the model checking processes that we instantiate, do not partition or parallelize verification computation, as our domain is different -- we seek to evaluate properties that otherwise devices in an IoT would compute, given the particularities of spatial verification. 
To the best of our knowledge, spatial model checking within our cloud context has not been considered before; spatial verification exhibits particular kinds of computational workloads, as we observed in Section~\ref{sec:spatialverifworkload}. Moreover, the potential offered to engineer applications in novel domains such as the IoT, has not been exploited, as discussed in the following.

We position our approach within engineering of dependable systems operating in a discrete space arising from topological relations in the spatial environment, where the information abstraction of physical \emph{location} or position of entities is inherently important~\cite{tsigkanos2016architecting,Lee.TechRep.2008}.
Topological relations have found application in various domains, where formal reasoning has proved beneficial, such as in foundational approaches~\cite{ciancia2014specifying,ciancia2016tool}.
In previous work~\cite{fse17} we 
considered spatio-temporal model checking of evolving spaces and
advocated that the topology can provide a system with awareness of multiple characteristics~\cite{tdsc,Tsigkanos.RE.2014,tsigkanos2016architecting}.
Closure space-based reasoning with the SLCS logic has been considered for data correctness in vehicle location data, in the context of collective adaptive systems~\cite{ciancia2014data}.
Moreover, a combination of CTL and SLCS is developed~\cite{ciancia2015exploring} to study bike sharing systems. 
In~\cite{sun2015specifying}, a combination of metric and spatial logics has been proposed for verification of safety properties in cyber-physical systems, while run time verification of spatio-temporal behaviours of complex systems is studied in~\cite{nenzi2015qualitative}, extending Signal Spatio-Temporal Logic with SLCS. 
For our evaluation purposes, we utilize a trajectory dataset, so we succinctly discuss relevant approaches. Trajectory-based reasoning has been studied extensively in scientific literature, where trajectories are typically modeled as streams of spatio-temporal points.
Recently the focus has been on the use of semantic abstractions of raw mobility data, including not only geometric information but also knowledge extracted jointly from the mobility data and the application domains information~\cite{yan2013semantic}. We note advanced data mining techniques which tackle related problems of low sampling-rate and uncertain features of trajectory data~\cite{chen2011clustering}, their context~\cite{wachowicz2013tailoring}, or forecasting~\cite{yan2010traj}.

%% file: sections/conclusions.tex

\section{Conclusions}
\label{sec:conclusion}

In this paper, motivated by the need for requirements assurance of spatially-distributed 
IoT systems, we proposed a technical framework for their spatial formal verification at runtime, leveraging a microservices architecture for the cloud.
 We defined the underlying model of space as a topological graph structure paving the way for expressing and verifying spatial logic properties. We devised a microservices architecture where we instantiated spatial model checking processes in a service layer, and
proposed alternative cloud deployments; Virtual Machines, containers and the recent Functions-as-a-Service. After investigating their tradeoffs when evaluating global spatial properties at runtime, we proposed a novel hybrid deployment, where preallocated resources are augmented with an additional FaaS fallback infrastructure serving unexpected peaks in the workload. Our evaluation with a 
realistic workload scenario investigates feasibility, performance, scalability and elasticity of different deployments.

Regarding future work, we aim to first challenge analysis assumptions of IoT-cloud coupling as well as enable the specification and verification of more complex spatial properties.
Today's IoT-cloud architectures assume that cloud resources are always available, and where software services then provide the analyses, business logic and data for devices to take action.
However, novel requirements driven by technological advancements such as timeliness, locality of computation or privacy may dictate analysis to take place closer to IoT devices, at the \emph{edge} of the network~\cite{glikson2017deviceless}.
For example, the round-trip network latency for spatial analysis (for smaller spatial models) may be prohibitive when analysis is performed on the cloud~\cite{GarrigaMendonca2017}, or fault-tolerance requirements may suggest that the central point of failure that the IoT-cloud coupling exhibits is not acceptable. As such, we aim to support analyses on the \emph{edge}~\cite{mendonca18toit} where the challenge is decentralizing verification as well as the appropriate data to support it, putting more focus on edge devices~\cite{bonomi2012fog} such as IoT gateways, network devices, cloudlets, and small clouds.
On the analysis side, we have demonstrated that relevant qualitative spatial requirements may be expressed with the SLCS logic. We plan to extend this predication to include quantitative~\cite{nenzi2015qualitative} and metric~\cite{sun2015specifying} aspects.
